\documentclass[11pt,a4]{article}
\usepackage{times,fancyhdr}
\usepackage[dvips]{graphicx}
\usepackage{natbib}
\usepackage{color} 
\usepackage{verbatim,framed}
\usepackage{amssymb}
\usepackage[footnotesize,bf]{caption}
\usepackage{setspace}

\definecolor{orange}{rgb}{1.0,0.5,0.0}
\definecolor{aqgr}  {rgb}{0.0,1.0,0.6} 
\definecolor{viol}  {rgb}{0.8,0.6,0.8}
\definecolor{figdr} {rgb}{1.0,1.0,1.0} 
\definecolor{coldr} {rgb}{1.0,0.0,1.0} 
\definecolor{colwp} {rgb}{1.0,0.8,0.0} 
\definecolor{colok} {rgb}{0.8,1.0,0.5} 

\title{\bfseries{Epigenetic Tracking: \\
a Model for all Biology}}

\author{Alessandro Fontana$^{1}$ \\
\mbox{}\\
$^1$IEEE \\
email address: cnd12001@yahoo.it}

\begin{document}
\maketitle


\begin{abstract}
``Epigenetic Tracking'' is a model of systems of biological cells, able to generate arbitrary 2 or 3-dimensional cellular shapes of any kind and complexity (in terms of number of cells, number of colours, etc.) starting from a single cell. If the complexity of such structures is interpreted as a metaphor for the complexity of biological structures, we can conclude that this model has the potential to generate the complexity typical of living beings. It can be shown how the model is able to reproduce a simplified version of key biological phenomena such as development, the presence of ``junk DNA'', the phenomenon of ageing and the process of carcinogenesis. The model links properties and behaviour of genes and cells to properties and behaviour of the organism, describing and interpreting the said phenomena with a unified framework: for this reason, we think it can be proposed as a model for all biology. The material contained in this work is not new: the model and its implications have all been described in previous works from a computer-science point of view. This work has two objectives: 1) To present the whole theory in an organic and structured way and 2) To introduce Epigenetic Tracking from a biological perspective. The work is divided into six parts: the first part is the introduction; the second part describes the cellular model; the third part is dedicated to the evo-devo method and transposable-elements; the fourth part deals with junk DNA and ageing; the fifth part explores the topic of cancer; the sixth part draws the conclusions. 
\end{abstract}

\pagebreak


\tableofcontents

\pagebreak

\section{Introduction}

\subsection{Embryogenesis and Artificial Embryology}

Embryogenesis, the process by which the zygote develops into a progressively more complex embryo to become an adult organism, is one of the greatest miracles of Nature, yet poorly understood. Such process is known to be guided by the DNA contained in the zygote, which is copied unaltered at each cell division. How does the DNA organise the growth process? Since all cells have exactly the same genetic makeup, how can each cell know which type of specialised cell it is destined to become? How are 3-dimensional structures generated from a linear DNA code? How do cells coordinate themselves to behave like an organism? 

Artificial Embryology is a sub-discipline of Artificial Life aimed at modelling the process of morphogenesis and cellular differentiation that drives embryogenesis. Models in the field of Artificial Embryology \cite{AY03KB} can be divided into two broad categories: grammatical models and cell chemistry models. In the grammatical approach development is guided by sets of grammatical rewrite rules; a prototypical example of grammatical models is represented by L-systems, first introduced by Lindenmayer \citep{AY68LX} to describe the complex fractal patterns observed in the structure of trees. The cell chemistry approach draws inspiration from the early work of Turing \cite{AY52TX}, who introduced reaction and diffusion equations to explain the striped patterns observed in nature; this approach attempts to simulating cell biology at a deeper level, reconstructing the networks of chemical signals exchanged within and between cells.

\subsection{Approach followed in constructing the model} 

This work is concerned with a cellular model stemming from the field of Artificial Embryology called ``Epigenetic Tracking'' (described in \cite{AX08AX}, \cite{AX09AX}, \cite{AX10A0}, \cite{AX10A1}, \cite{AX10A2}), able to generate arbitrary 2- or 3-dimensional cellular structures starting from a single cell. The model has been constructed with the following approach: first, we have designed the architecture of the model at a high level, based on known biological elements. Subsequently, we have added additional elements -not necessarily known- in order to ``make things work in silico'' (i.e. to produce interesting behaviours through computer simulations). Finally, we have come back to biology, trying to guess which biological molecules play the role of the additional elements. As a consequence the model contains ingredients not necessarily adherent to current knowledge, but which can become a suggestion for biologists to look into new, previously unexplored directions.
 
Computer simulations have proved the capacity of the model to generate any kind of shape, of any complexity (in terms of number of cells, number of colours, etc.) starting from a single cell. If the complexity of such structures is interpreted as a metaphor for the complexity of biological structures, we can conclude that this model has the potential to generate the complexity typical of living beings. Furthermore, it can be shown how the model is able to reproduce a simplified version of key biological phenomena such as development, the presence of ``junk DNA'', the phenomenon of ageing and the process of carcinogenesis. To our knowledge, this is the only model able to describe and interpret the said phenomena with a unified framework: for this reason, we think it can be proposed as a model for multicellular biology.
 
The material contained in this work is not new: the model and its implications have all been described in previous works from a computer-science point of view. This work has two objectives: 1) To present the whole theory in an organic and structured way and 2) To introduce Epigenetic Tracking from a biological perspective. In order to more closely reflect the biological interpretation, the terminology has been changed compared to previous papers. This work is divided into six parts: this first part is the introduction; the second part describes the cellular model; the third part is dedicated to the evo-devo method and transposable-elements; the fourth part deals with junk DNA and ageing; the fifth part explores the topic of cancer; the sixth part draws the conclusions and outlines future directions.

\pagebreak[4]

\section{The cellular model of development}

\subsection{Driver cells, the genome and the epigenome}

In this model phenotypes are represented as {\em cellular shapes}, aggregates of cube-shaped cells deployed on a grid. Cells belong to two distinct categories: ``normal'' cells, which make up the bulk of the shape and ``driver'' cells, which are much fewer in number (by orders of magnitude) and are evenly distributed in the shape volume. If we compare normal cells to simple soldiers of an army, driver cells can be compared to sergeants; in other words, driver cells have the power to issue orders that normal cells have to obey. As we will see, this is the single most important feature of the model.

Development (see \colorbox{figdr}{figure 1}) starts with a single cell (called zygote) placed in the middle of the grid and unfolds in N (developmental) steps, counted by a ``global clock'' (GC), shared by all cells.  During development, a shape can be ``viewed'' in two ways: in ``external view'' colours are used to represent cell types; in ``internal view'' colours are used to represent cell states: blue is used for normal cells alive, orange for normal cells just created (i.e. in the current step), grey for cells that have just died, yellow for driver cells (regardless of when they have been created).

The genetic information (\colorbox{figdr}{figure 2}), encoded as in nature by sequences of four numbers (0,1,2,3), is composed of two sets:
\begin{itemize}
\item the ``fixed'' genome and
\item the ``variable'' genome or epigenome.
\end{itemize}
The fixed genome, as the name implies, is equal in all cells; the variable genome, on the other hand, can be different in different cells, reflecting the differentiation occurring during development. 

The fixed genome is structured as an array of ``genes'': each gene is composed of a left (or ``if'') part, encoding a condition, and a right (or ``then'') part, encoding an action: once the condition is verified, the action is executed. There are two types of genes:
\begin{itemize}
\item metabolic genes (marked in brown), which are responsible for cellular behaviour at steady state, and
\item developmental genes (marked in yellow), which are responsible for the changes occurring to the organism.
\end{itemize}

The variable genome, as we said, is composed of the elements which can become different in different cells. These elements can be grouped in four categories:
\begin{itemize}
\item methylation marks, which control the activation of metabolic genes; 
\item mobile elements, which are placed in precise genomic loci, and render developmental genes active; 
\item the Mobile Code (MOC), a set of mobile elements stored in a separate cellular compartment;
\item the driver status, indicating whether a cell is driver or not.
\end{itemize}
 
An esoteric element of the variable genome is the MOC, present in both normal and driver cells, but {\em active} in driver cells only. MOC sequences (indicated with letters: A, B, C, etc.) are changed during development, so that they become {\em different} in each driver cell. The MOC is used by different driver cells as a ``key'' to activate different portions of the genome: in this way it represents the source of differentiation during development (\colorbox{figdr}{figure 3}). The presence of a piece of genetic material different in each (driver) cell marks a fundamental difference with respect to other models of development and also with respect to current knowledge.  
 
Metabolic genes have a structure which closely resemble that of real genes. The left part of a metabolic gene (\colorbox{figdr}{figure 4}) is composed of the following elements: a field called ``switch'' (SWC), indicating whether the gene is ``structurally'' active (1) or not (0); two sequences called ``transcription factor binding sites'' (TFS), where matching transcription factors can bind. The right part of the gene is composed of a single element, called ``output transcription factor'' (OTF), destined to become a transcription factor for other metabolic genes. Taken together, metabolic genes compose a gene regulatory network which, thanks to the selective activation / deactivation of individual genes, becomes different in different cells, allowing specialisation to take place. The selective activation / deactivation of metabolic genes is done changing the value of their methylation mark, an action  carried out by developmental genes.
 
Developmental genes can be compared to ``macro'' genes or to sets of genes co-regulated. The left part of a developmental gene (\colorbox{figdr}{figure 4}) is composed of the following elements:
\begin{itemize}
\item a field called ``switch'' (SWC), indicating whether the gene is ``structurally'' active (1) or not (0);
\item a sequence called ``mobile insertion site'' (MOS), than can match with the MOC;
\item a sequence called ``timer'' (TM), than can match with GC (the clock).
\end{itemize}

At each developmental step (marked by a different clock value), for each developmental gene and for each driver cell, it is checked if the gene's MOS matches the driver's MOC and if the gene's TM matches GC (MOC and MOS sequences are indicated with letters: if a MOC and a MOS are labelled with the same letter, it means they are equal). If both conditions are verified, the right part of the gene is executed.
 
In practise, the MOC plays the role of a mobile genetic element which, upon binding to, or inserting into its genomic target site (the MOS), activates a developmental gene, triggering the action encoded by the relevant right part. The lasting effect of such insertions is represented by mobile elements permanently placed in the developmental genome (positions marked in green in \colorbox{figdr}{figure 4}).

TM behaves like a timer, imposing a further condition on the activation of the gene, namely that the clock reaches a certain value. In a possible biological implementation, the clock value is physically represented by a protein, which enters the cell and is transduced into the nucleus by means of a dedicated piece of cellular machinery. The transduced molecule binds to the genomic sequence corresponding to the timer, fulfilling the temporal condition for the activation of the gene.  
 
The right part of a developmental gene has four fields:
\begin{itemize}
\item a field called ``metabolic gene activation / deactivation'' (MAD);
\item a field called ``change event type'' (EVT);
\item a field called ``shape'' (SHP);
\item a field called ``colour'' (COL).
\end{itemize}

The field MAD influences the switch of some metabolic genes, turning them on or off. The field EVT encodes the type of ``change event'' (\colorbox{figdr}{figure 5}): ``proliferation events'' cause the activated driver cell (called ``mother driver cell'' or simply ``mother cell'') to proliferate in the volume around it (called ``change volume''); ``apoptosis events'' cause cells in the change volume to be deleted from the grid. The field SHP specifies the shape of the change volume, in which the event takes place. In case of proliferation, the field COL specifies the colour of the newly created cells. In case of apoptosis, the fields MAD and COL have no effect.
 
We can imagine that a proliferation event starts by producing a pool of undifferentiated, stem-like cells, based on the footprint of the driver cell activated (under the control of parameter SHP); these cells are then induced to differentiate (under the control of parameters MAD and COL). This process could be implemented in nature through a class of molecules called {\bf growth factors}, which during embryonic development act locally, either as paracrine or autocrine regulatory chemical messengers, as important regulators of cellular proliferation and differentiation \cite{EA09RD}.

In case of proliferation, it may happen that the change volume is not empty. In this case the cells present in the change volume must be either deleted or moved to other positions, to make room for the cells that are going to be  created. This correponds to implementing a model of ``physics'', which can be defined as the set of rules by which cells are moved around in the shape and find their final position. Many models of physics can be implemented, based on the trade-off between degree of realism and computational burden, as long as the deterministic requirement is maintained, i.e. as long as physics behaves in way that is predictable and consistent, as we all expect.

\subsection{Generation of new driver cells}

During a proliferation event, initially only normal cells are created in the change volume (see \colorbox{figdr}{figure 6}). On the other hand, the presence of a uniform distribution of driver cells in the shape volume throughout development is a key feature of this model. Indeed, if a part of the shape remained deprived of driver cells, no change event could occur in that part and further development would be impeded. Therefore new driver cells need to be created in the change volume. A further requirements is that each driver cell must have a distinct MOC sequence, as otherwise all driver cells with the same MOC sequence would be activated by the same developmental gene. This would represent a constraint that must be avoided.
 
This is achieved by means of the following mechanism. Each (existing) driver cell (see \colorbox{figdr}{figure 6}) continuously emits in the neighbourhood a mix of chemicals characteristic for each driver cell (the mix emitted by a driver cell is {\em locally} different from the mix emitted by any other driver cell). Each normal cell keeps monitoring the concentrations of such chemicals; if the strongest concentration signal received falls below a given threshold (meaning that the closest driver cell is not close enough), the cell ``decides'' to turn itself into a driver cell. As the new driver cell starts diffusing its own chemical fingerprint, nearby cells are automatically inhibited from undergoing the same transformation, and the system reaches a state of equilibrium, characterised by a typical ratio between driver and normal cells.
 
To each new driver cell a new and unique MOC sequence must be assigned. This is achieved through the transduction of the chemical signals released by the surrounding drivers (see \colorbox{figdr}{figure 7}) into the compartment where the MOC is stored of the normal cell destined to become a driver (the MOC is currently inactive as this used to be a normal cell). Such chemicals are processed by a dedicated piece of cellular machinery (the red box in the figure), which takes as inputs the chemicals and produces a new sub-sequence which is appended at the end of the existing MOC sequence to produce the new MOC sequence for the driver cell. As we said, this mechanism has to ensure that the new sequences are unique.
 
It is here conjectured that these chemicals, released by driver cells and processed by the piece of cellular machinery which produces the MOC sequence, are implemented in nature by a class of chemical compounds called {\bf morphogens}. Morphogens are substances involved in the patterning of tissue development and the positioning of the various specialised cell types within a tissue. They are spread from a localised source and provide spatial information by forming a concentration gradient and inducing the expression of specific genes at distinct concentration thresholds. Different combinations of morphogens induce the emergence of new cell types through a characteristic ``morphogenetic code'' \cite{EA99HX}. An analogous role could be played by different combinations of the products of HOX genes, implementing a ``HOX code'' \cite{EA91KG}.
  
In summary, during a proliferation event (\colorbox{figdr}{figure 8}): new normal (undifferentiated) cells are created (B) and sent down a differentiation path (C), represented by the colour and by the specialised metabolic network; in parallel other driver cells are generated, each with its own MOC sequence, one of which could subsequently become the centre of another event of proliferation or apoptosis. \colorbox{figdr}{figure 9} shows an example of development in three steps. The right part of the figure reports the set of MOC sequences created during this development. Since each driver cell is the result of a conversion from a normal cell, generated by a single mother driver cell, MOC sequences are organised in a hierarchical tree-like structure.   

Each driver cell generated during development (see \colorbox{figdr}{figure 3}) will have a set of MOC elements bound to their corresponding MOS sequences in developmental genes (reflecting the record of all developmental genes activated during development in the lineage of that cell), and a specialised metabolic network (reflecting the record of the activations / deactivations carried out by all developmental genes activated during development in the lineage of that cell).

\subsection{Biological interpretation of driver cells}

A key element in the described cellular model is represented by driver cells, a subpopulation of cells which ``drive'' development. Only driver cells can be instructed to develop (proliferate or undergo apoptosis) directly by the genome, while normal cells need to get such instructions from driver cells. This makes it possible to steer development by acting on a small subset of cells, representing the scaffold and the backbone of the developing shape. The MOC sequence stored inside the driver cell is different in each driver cell and represents the source of differentiation during development. This feature marks a fundamental difference with respect to other models of development, which rely on the positional information and the chemical micro-environment to provide the information necessary for differentiation.
 
In natural development a key role is known to be played by stem cells, a class of cells found in most multi-cellular organisms. {\bf Embryonic stem (ES) cells} (found in the inner cell mass of the blastocyst) are totipotent cells, which means that they are able to differentiate into all cell types of the body. {\bf Adult stem cells} are pluripotent undifferentiated cells found throughout the body after embryonic development. Unlike totipotent ES cells, adult stem cells can only form a limited set of cell types and function to replenish dying cells and regenerate damaged tissues.
 
A spontaneously arising question regards the natural counterpart of driver cells and its relation to stem cells. As we said, driver cells can be compared to the sergeants of an army in which normal cells are the simple soldiers: in other words, they have the power to induce normal cells to perform certain actions. The key question here is whether stem cells are cells which induce other, ``normal'' cells, to undertake actions and commit to specific fates or, conversely, they are induced by other cells to undertake actions, being their susceptibility to be induced proportional to their degree of ``plasticity''.
 
Many analogies exist between the concept of driver cell and the concept of Spemann's organiser, which is an area of the Xenopus embryo able to induce embryonic primordia upon transplantation into a different location \cite{EA06DR}. Likewise, if a driver cell destined to give rise to a certain shape part is moved to a different position of the growing shape, such shape part will grow in the new, ectopic position. 

In our model a single driver cell can produce a body structure in a fully autonomous way. When induced to proliferate, such driver cell will itself induce the creation of new driver cells which, in turn, will become the centres of other proliferation or apoptosis event, from which other driver cells will be created and so on, until the whole structure is generated. This appears to be consistent with the so-called ``head / trunk / tail organiser model'' \cite{EA06SX}, which foresees the presence of many organisers. In our model the organisers (the driver cells) are many, are hierarchically structured and are continuously created during development.
 
Driver cells drive development and are present for the whole duration of the organism life, from the zygote stage to the moment of death, a property that marks a fundamental difference with respect to current knowledge, and that will be shown to have profound biological implications. A special type of driver cells can be introduced to model tissue maintenance and repair. Such driver cells, when proliferating, generate both driver cells with new MOC sequences and driver cells with the same MOC sequence as their mother: we can call such driver cells ``maintenance'' driver cells.
 
Based on these considerations, we propose a new classification of biological cells (\colorbox{figdr}{figure 10}):
\begin{enumerate}
\item ``(natural) driver cells''. These are the natural counterpart of driver cells in the model and correspond to biological organisers. They can create and / or induce stem cells to undertake actions and commit to specific fates. Driver cells drive development, maintenance driver cells coordinate tissue repair; 
\item ``stem cells''. These are cells characterised by a high degree of plasticity and are susceptibile to be induced by driver cells to take different cellular fates. Embryonic stem cells are induced to differentiate by driver cells, adult stem cells by maintenance driver cells; 
\item ``normal cells''. These are cells with no plasticity, i.e. they are terminally differentiated cells. Actually we can imagine stem cells and normal cells as the extremes of a continuum of cells characterised by a decreasing degree of plasticity.   
\end{enumerate}
 
The key innovation in this scheme is the introduction, besides normal and stem cells, of a third layer of cells, namely natural driver cells, which represent the scaffold of the organism. During development some of them (those highlighted in green) become activated thanks to specific developmental genes. As a result of their activation, a local pool of stem cells, specific for the body part under construction, is produced and induced to differentiate into the specialised cell types needed, by means of growth factors. At the same time, other driver cells are generated and homogeneously distributed in the part just created, some of which are destined to become centres of other developmental acts in the course of development. The plasticity of stem cells depends on the driver cells that generates them and, in general, we can think of it as a decreasing function of developmental time.
 
A key ingredient of driver cells, the MOC, needs also to have a biological counterpart. The MOC is the ``code'' that distinguishes different driver cells and inserts itself into a corresponding MOS sequence in the left part of a developmental gene, from which a change event, such as proliferation or apoptosis, is produced. In the case of proliferation, besides the bulk of normal cells that are being sent down a differentiation path, other driver cells are created and assigned a corresponding number of different MOC sequences. These codes represent the ``handle'' by means of which such driver cells can be given other instructions to execute at a subsequent developmental step. From all we said it is clear that the MOC is an absolutely essential ingredient of Epigenetic Tracking, and must have a natural counterpart.
 
The hypothesis that will be proposed here is that the biological counterpart of the MOC in a natural driver cell is represented by a set of transposable elements (TE). Such TE's could be stored in an extrachromosomal compartment and, when needed, they would ``jump'' into precise locations on chromosomes, activating the natural counterpart of specific developmental genes (\colorbox{figdr}{figure 11}). A consequence of this hypothesis is that the TE configuration of a natural driver cell is not fixed, but is modified in the course of development; in this way different natural driver cells acquire different TE configurations which, like the MOC, represent the source of differentiation during development. As far as the nervous system is concerned, this view is supported by a recent work \cite{EA10SM}, which brings evidence that the copy number of LINE1 TE's is different in different nerve stem cells. 

\begin{figure}[p] \begin{center}
{\fboxrule=0.2mm\fboxsep=0mm\fbox{\includegraphics[width=11.50cm]{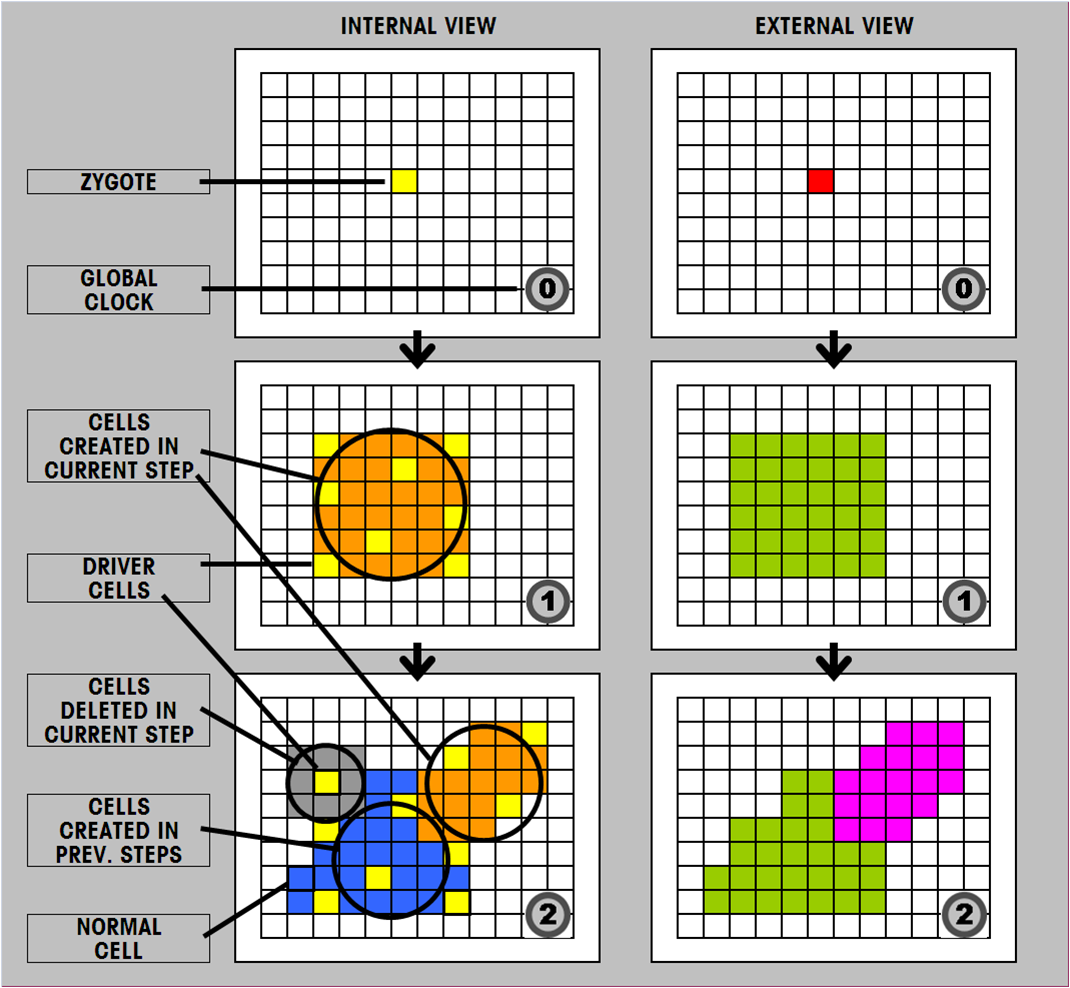}}}
\caption{Development and views. Development starts with a single cell and unfolds in N steps, counted by a ``global clock''.  During development, a shape can be seen from an internal and from an external perspective.}
\label{figxx}
\end{center} \end{figure}

\begin{figure}[p] \begin{center} \hspace*{-0.25cm}
{\fboxrule=0.2mm\fboxsep=0mm\fbox{\includegraphics[width=16.50cm]{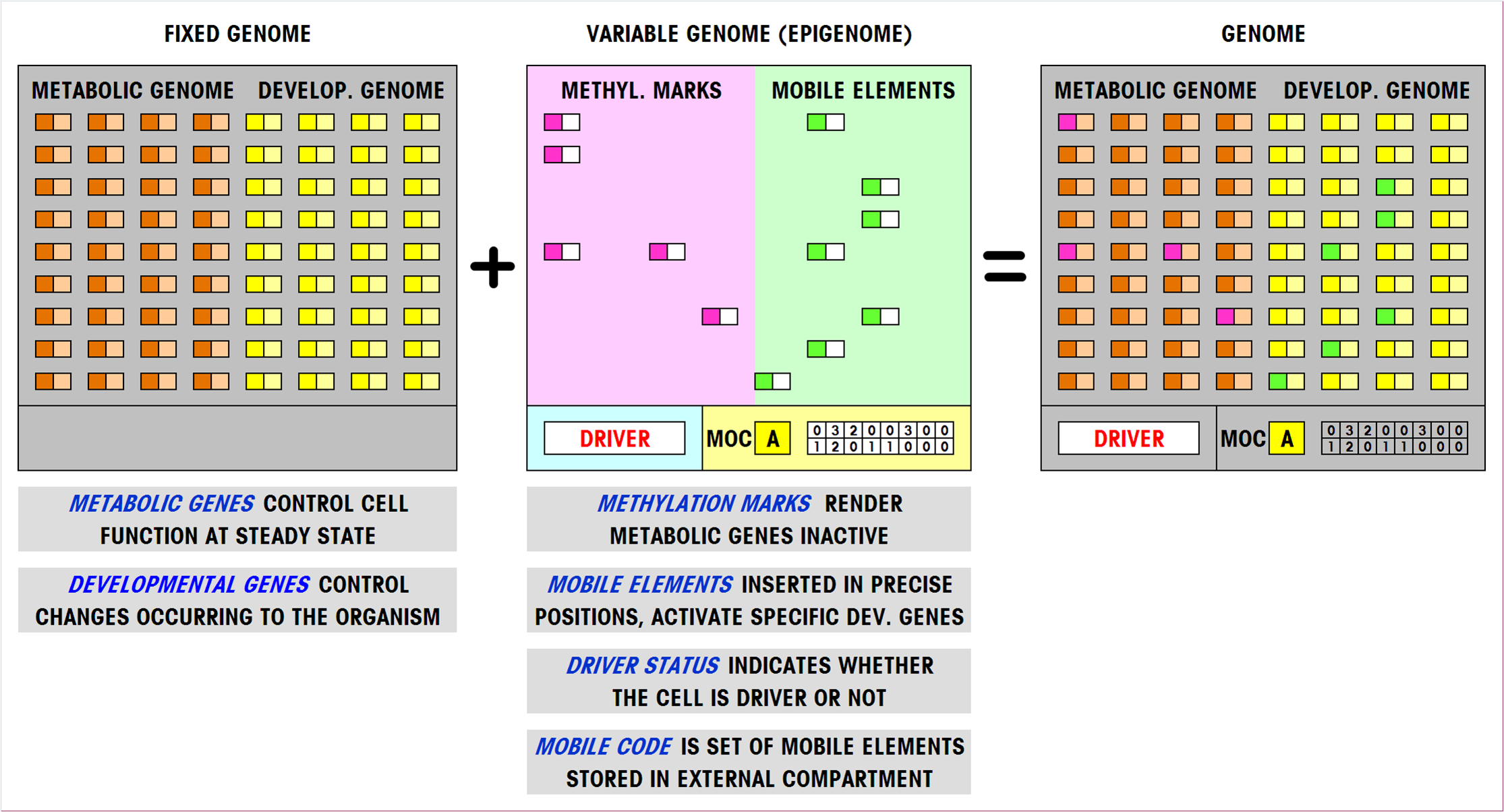}}}
\caption{Fixed and variable genome. The fixed genome is equal in all cells, while the variable genome becomes different in different cells, reflecting the differentiation occurring during development.}
\label{figxx}
\end{center} \end{figure}

\begin{figure}[p] \begin{center}
{\fboxrule=0.2mm\fboxsep=0mm\fbox{\includegraphics[height=09.00cm]{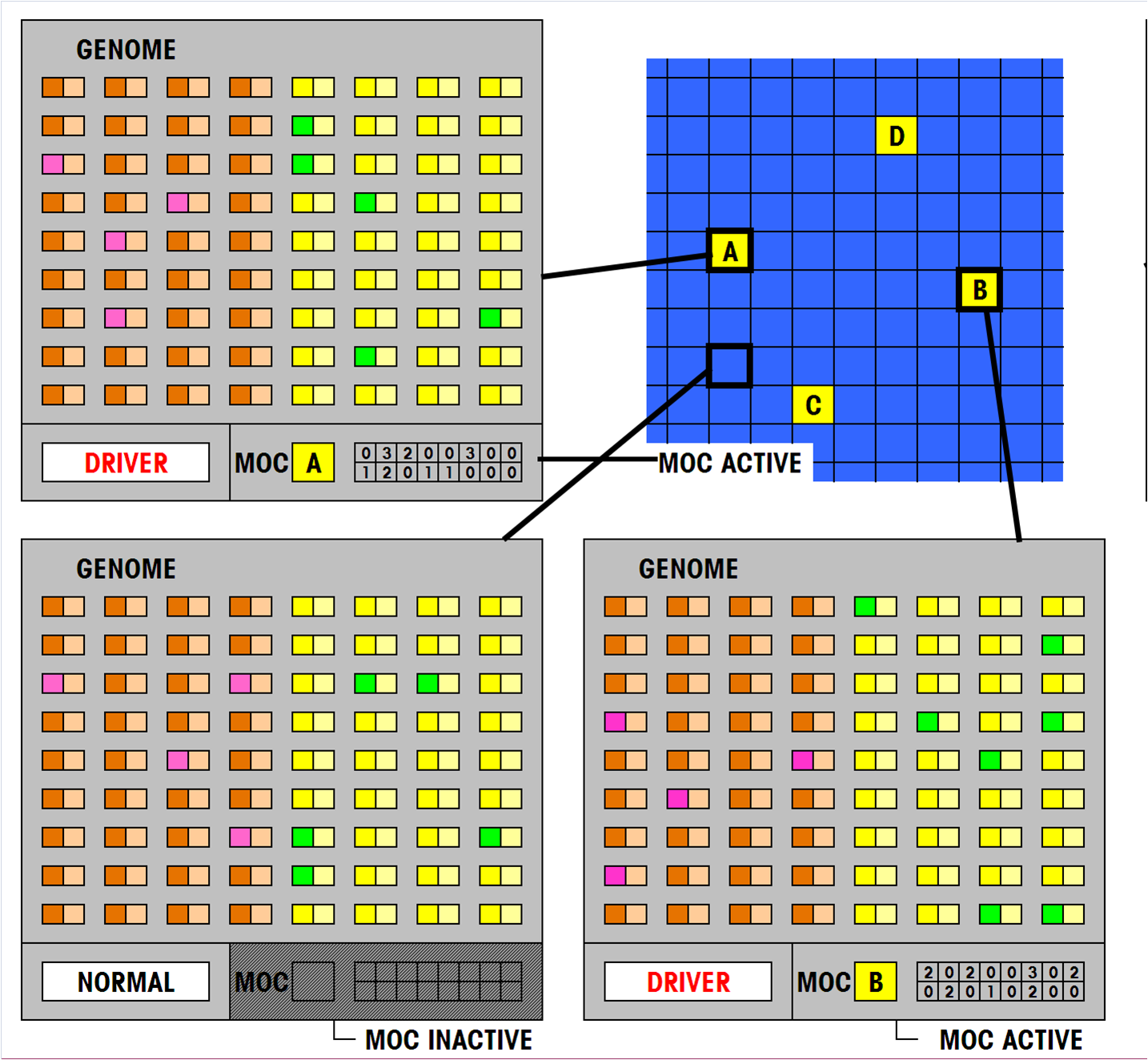}}}
\caption{Examples of differentiated cells. The differentiation is produced by the interplay of a differentiated epigenome and a constant fixed genome.}
\label{figxx}
\end{center} \end{figure}

\begin{figure}[p] \begin{center}
{\fboxrule=0.2mm\fboxsep=0mm\fbox{\includegraphics[width=12.00cm]{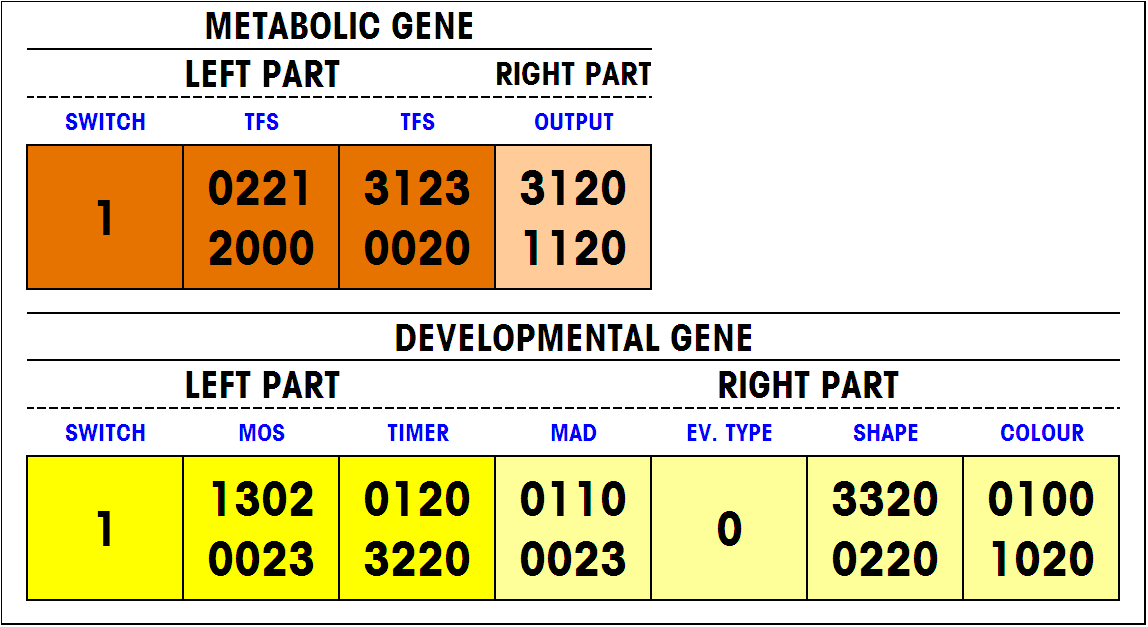}}}
\caption{The structure of a metabolic gene (top) and of a developmental gene (bottom).}
\label{figxx}
\end{center} \end{figure}

\begin{figure}[p] \begin{center}
{\fboxrule=0.2mm\fboxsep=0mm\fbox{\includegraphics[width=13.00cm]{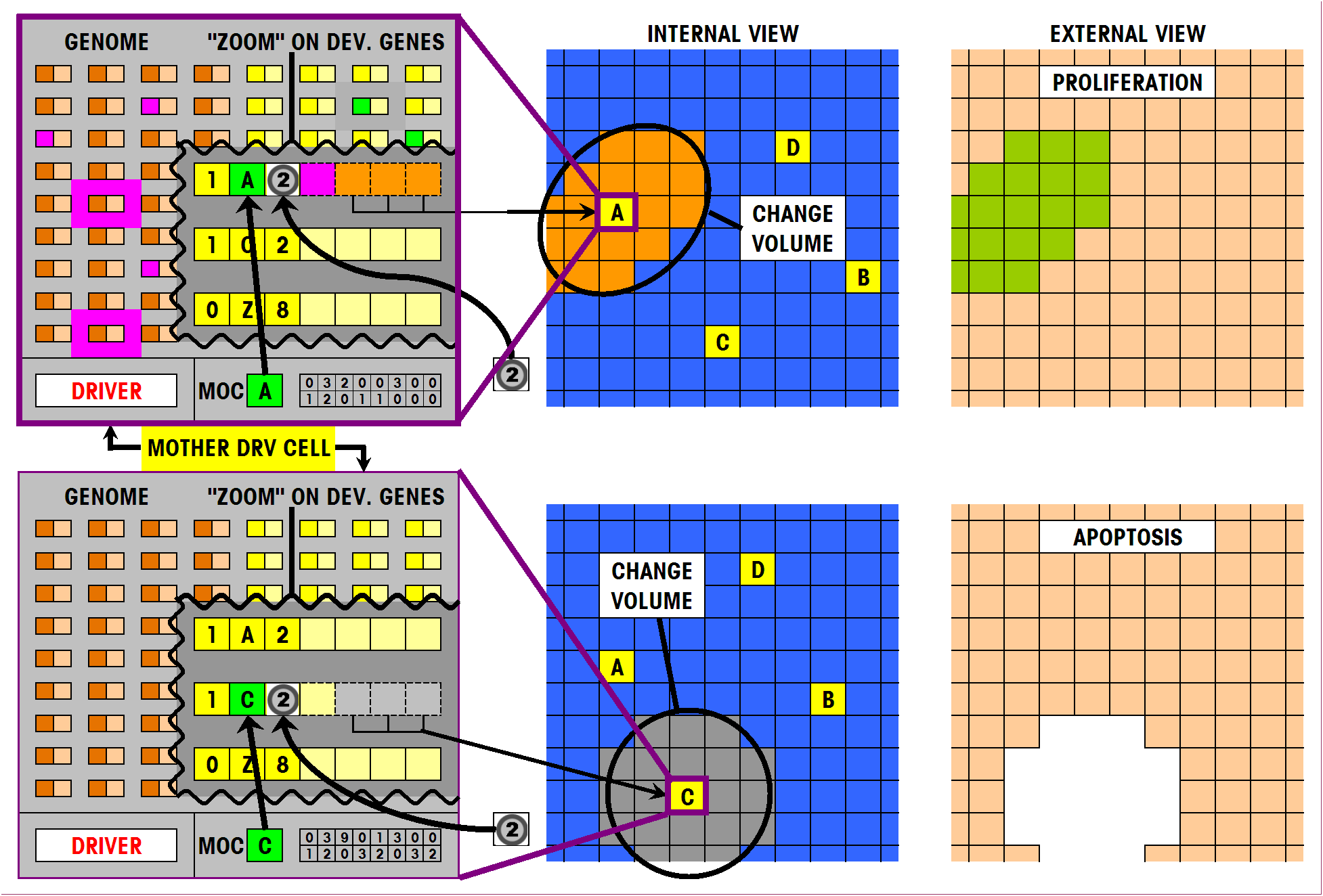}}}
\caption{In the upper part a change event of type proliferation. MOC sequence A matches with MOS sequence A and the clock value matches with the timer value (2): since the switch is 1, the gene is activated. The right part codes for a diagonal-shaped proliferation event. Before proliferation, the MAD field of the gene changes the switches of some metabolic genes (marked in purple), turning them on or off (the genes inactive are marked in purple). Such pattern of metabolic gene activation is inherited by all newly generated cells. In the lower part a change event of type apoptosis.}
\label{figxx}
\end{center} \end{figure}

\begin{figure}[p] \begin{center}
{\fboxrule=0.2mm\fboxsep=0mm\fbox{\includegraphics[width=11.00cm]{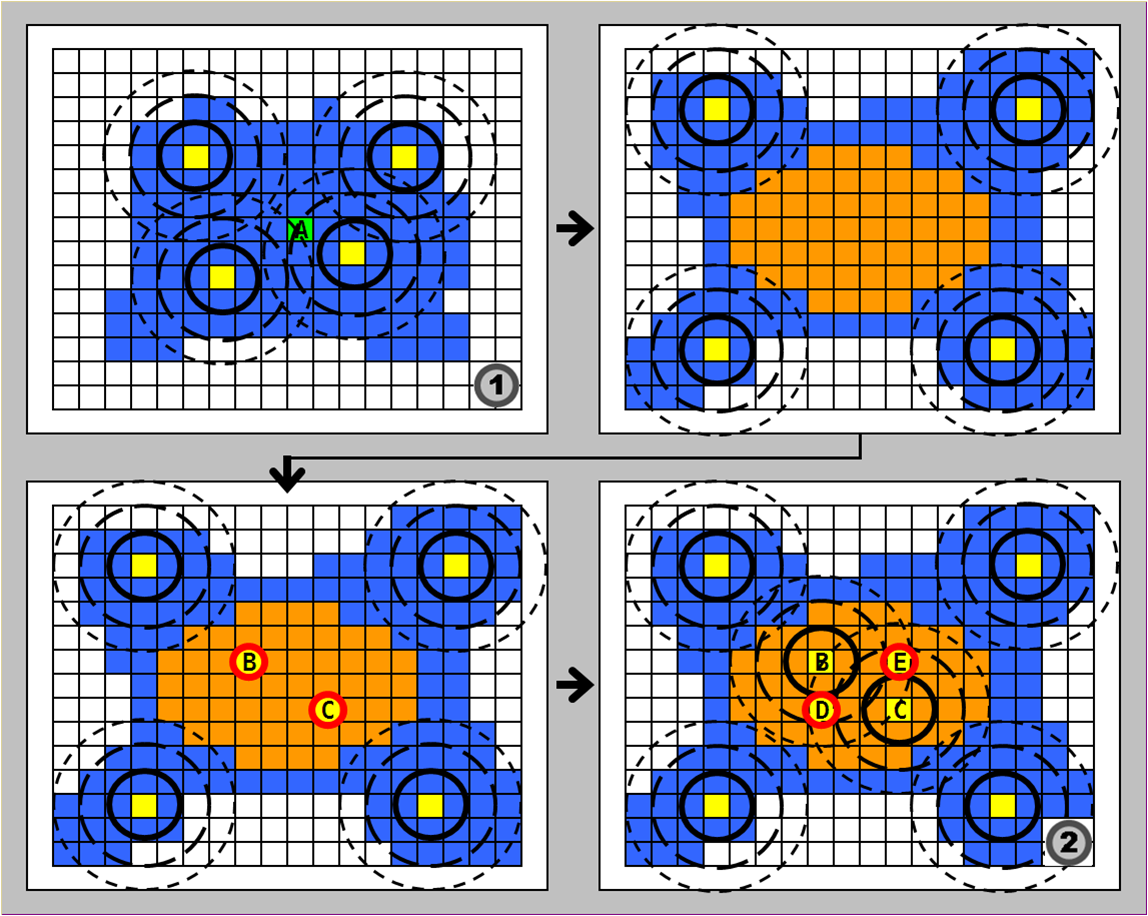}}}
\caption{Generation of new driver cells during a proliferation event. Driver cells continuously emit chemicals in the surrounding environment. Normal cells continuously monitor the surrounding chemical landscape: whenever a normal cell senses a too weak level of chemicals, it turns itself into a driver cell. As the newly created driver cells start emitting their own mix of chemicals, neighbouring cells are prevented from becoming driver.}
\label{figxx}
\end{center} \end{figure}

\begin{figure}[p] \begin{center}
{\fboxrule=0.2mm\fboxsep=0mm\fbox{\includegraphics[width=12.50cm]{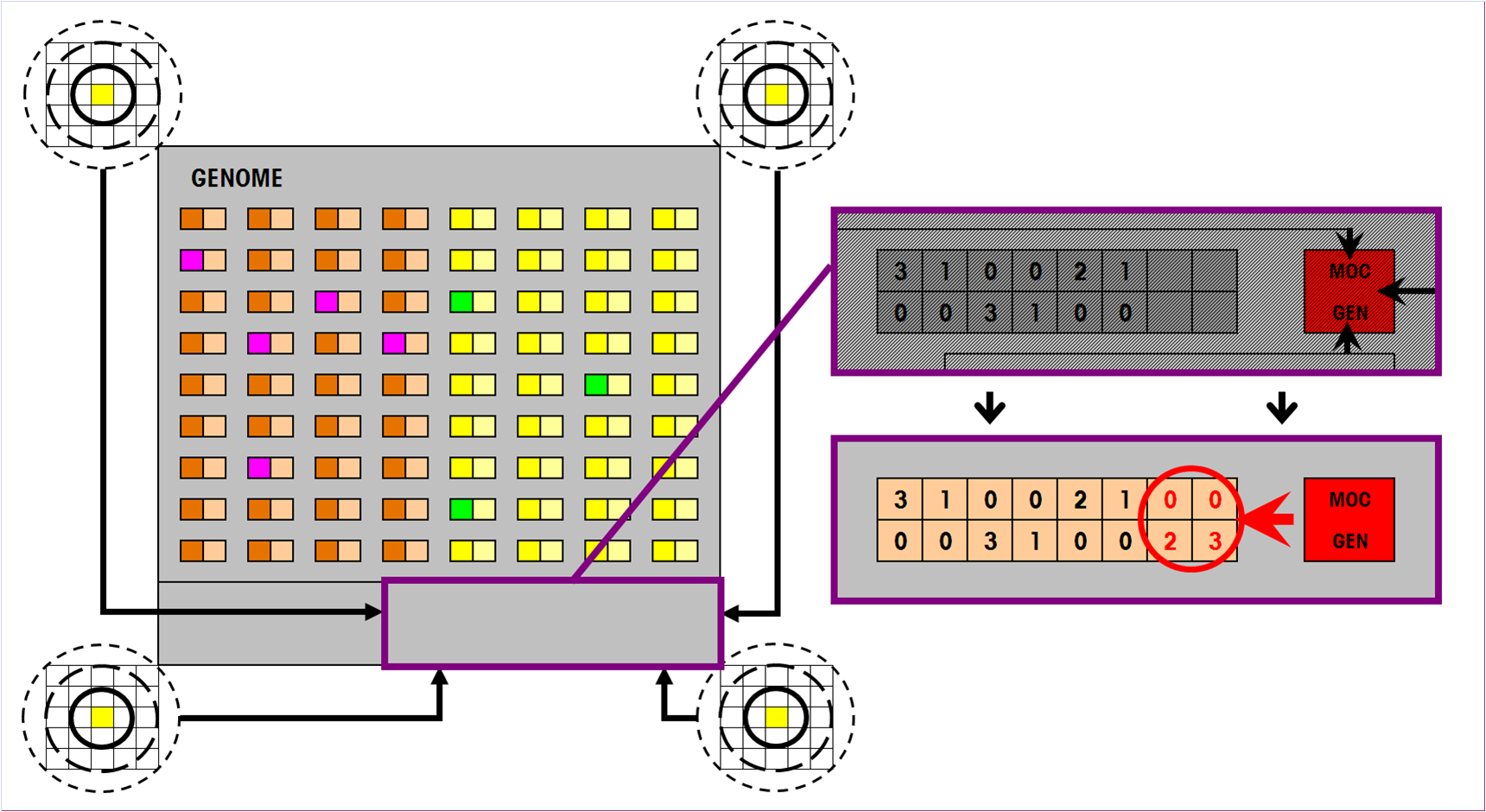}}}
\caption{Generation of new MOC sequences in new driver cells. The chemical signals emitted by the neighbouring driver cells reach a piece of cellular machinery called MOC generator (shown in red), that processes them and produces as output a new element of code which is appended to the Mobile Code inherited from the mother. The MOC is then activated, completing the process of creation of the new driver cell.}
\label{figxx}
\end{center} \end{figure}

\begin{figure}[p] \begin{center}
{\fboxrule=0.2mm\fboxsep=0mm\fbox{\includegraphics[height=10.00cm]{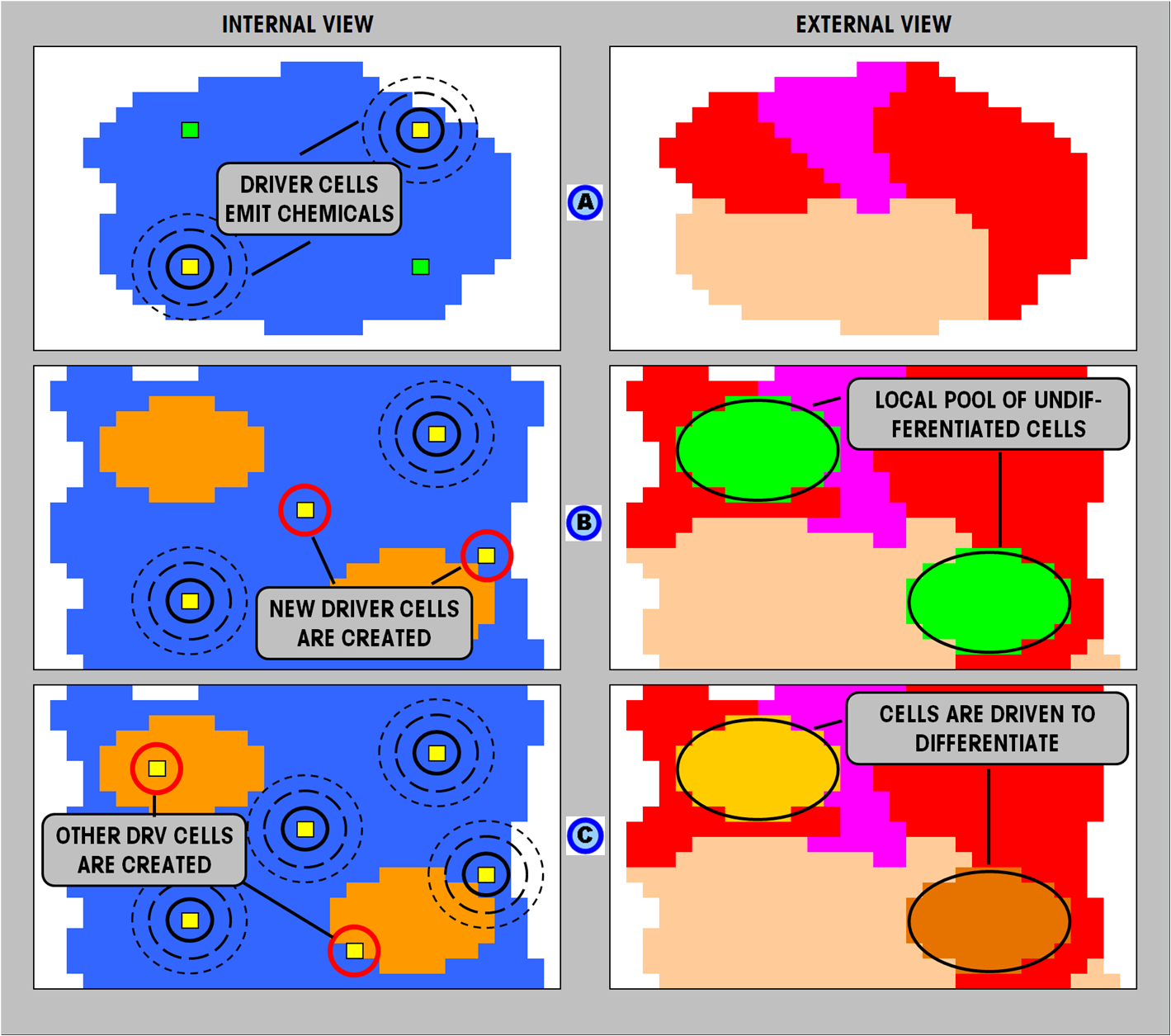}}}
\caption{During a proliferation event, new normal (undifferentiated) cells are created (B) and sent down a differentiation path (C), represented by the colour and by the specialised metabolic network; in parallel other driver cells are generated.}
\label{figxx}
\end{center} \end{figure}

\begin{figure}[p] \begin{center}
{\fboxrule=0.2mm\fboxsep=0mm\fbox{\includegraphics[width=13.50cm]{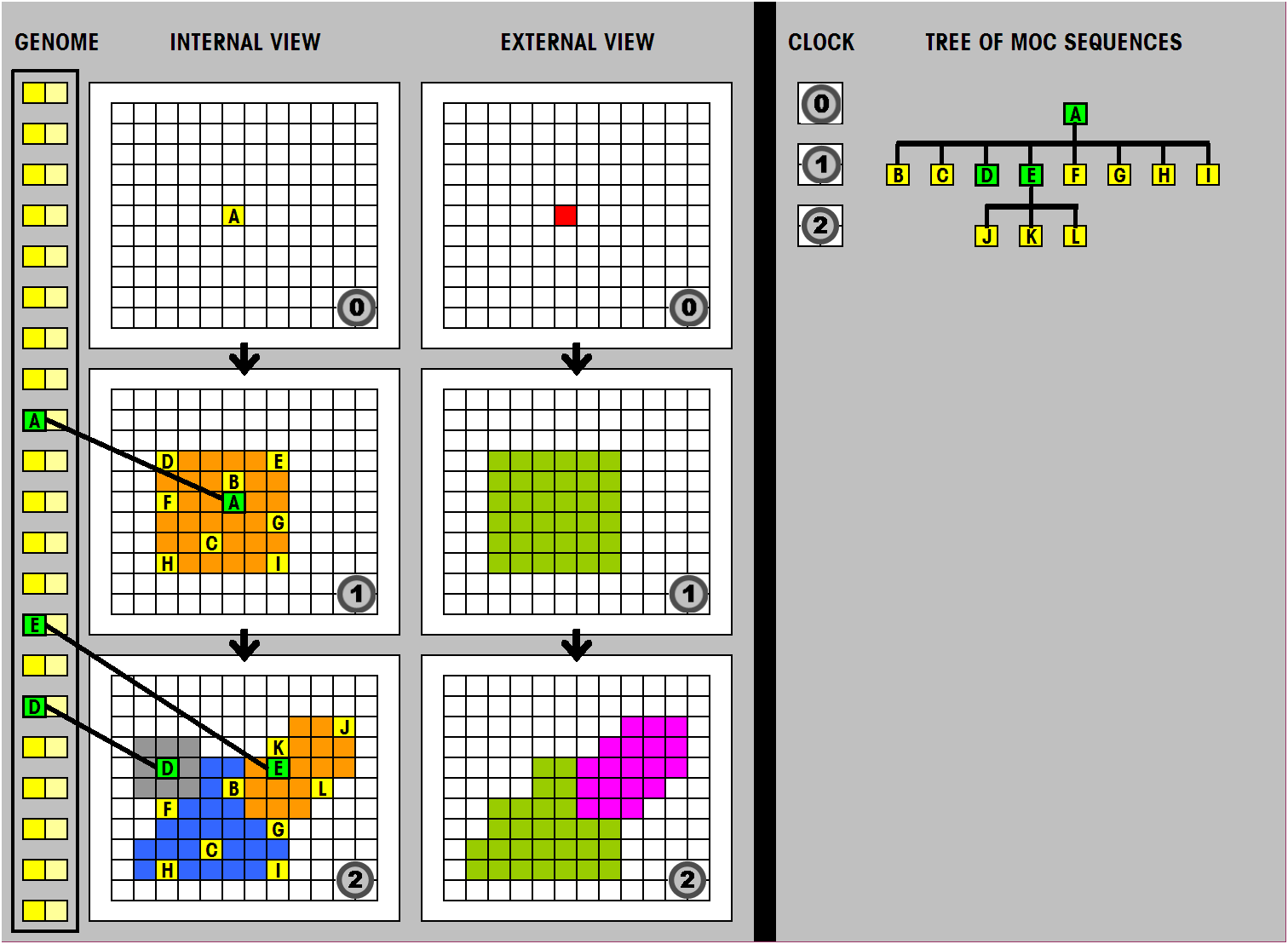}}}
\caption{Example of development in three steps (0,1,2): in step 1 the driver cell labelled with A is triggered to proliferate, in step 2 the driver cell labelled with D is triggered to undergo apoptosis and the driver cell labelled with E is triggered to proliferate. The developmental part of the genome is shown on the left of the figure, only once as it is common to all cells.}
\label{figxx}
\end{center} \end{figure}

\begin{figure}[p] \begin{center}
{\fboxrule=0.2mm\fboxsep=0mm\fbox{\includegraphics[height=07.00cm]{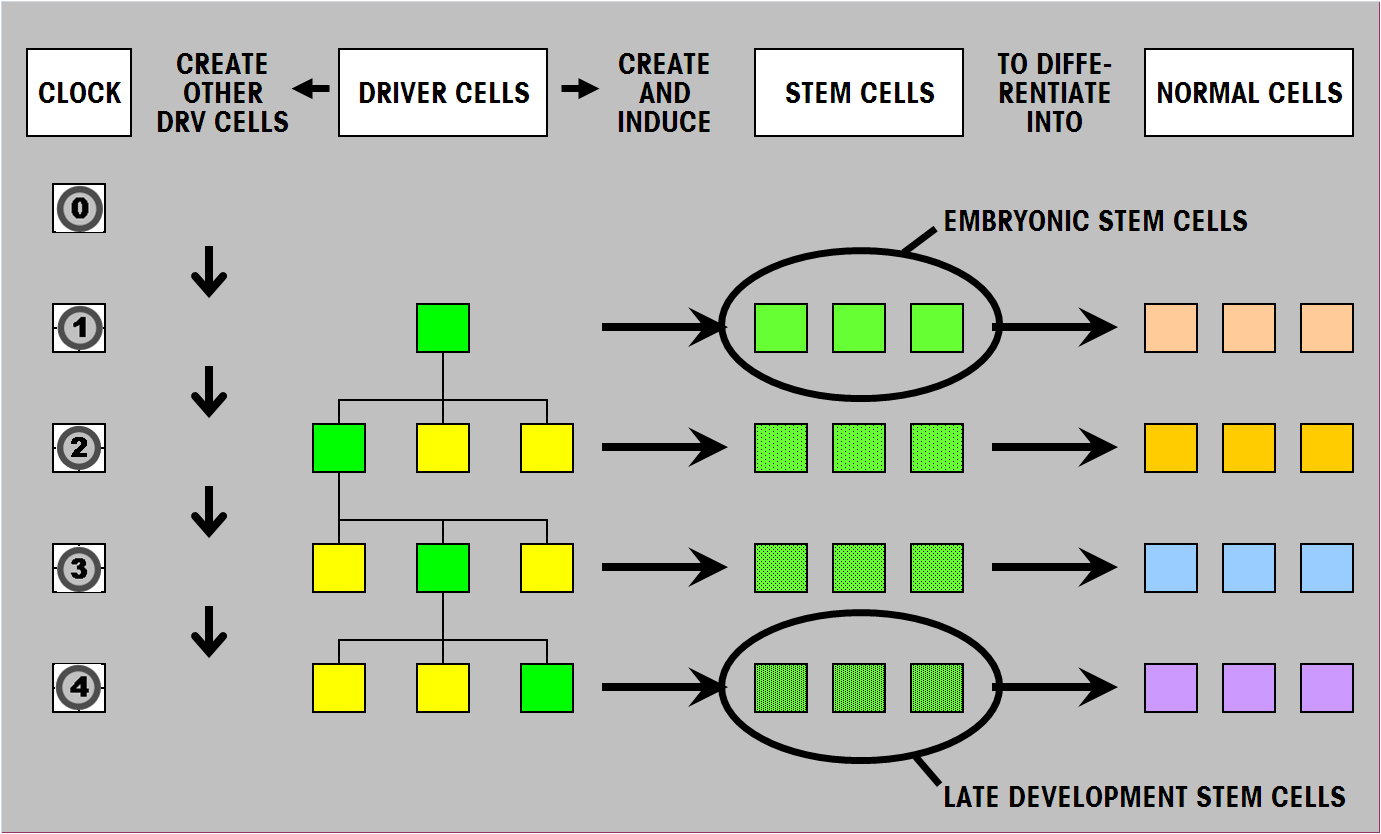}}}
\caption{Proposed classification of natural cells. Driver cells marked in green become active during development. As a result, they create and / or induce stem cells to differentiate into normal cells. Besides, driver cells create other driver cells. Normal cells can be induced to further differentiate, or trans-differentiate, or even de-differentiate by other driver cells at a subsequent step. ``Darker'' stem cells have a lower degree of plasticity.}
\label{figxx}
\end{center} \end{figure}

\begin{figure}[p] \begin{center}
{\fboxrule=0.2mm\fboxsep=0mm\fbox{\includegraphics[height=08.00cm]{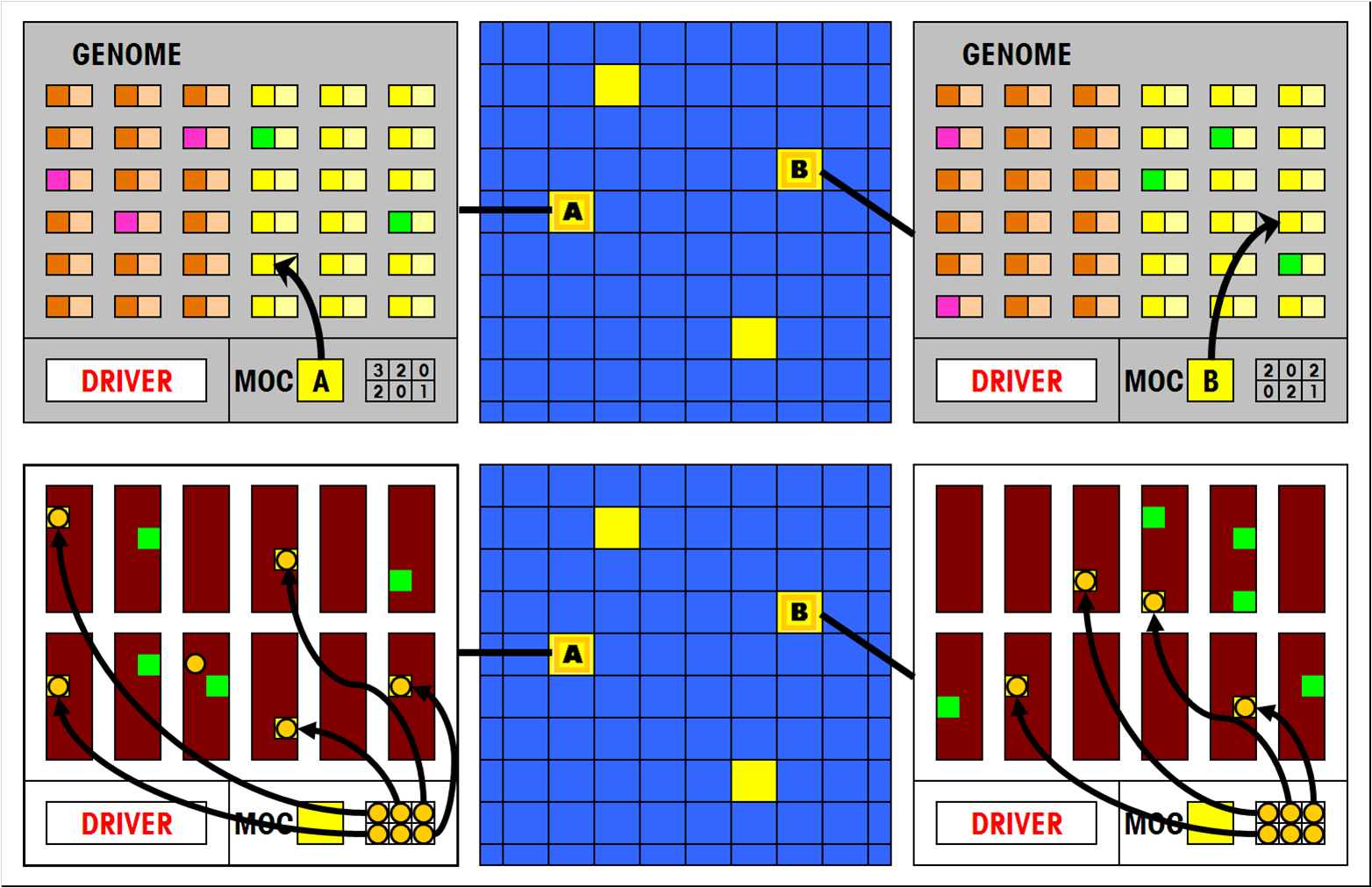}}}
\caption{In natural driver cells MOC sequences are implemented as sets of transposable elements stored in an extra-chromosomal compartment. Such transposable elements would jump into precise chromosomal locations, thereby activating / deactivating specific genes.}
\label{figxx}
\end{center} \end{figure}

\clearpage

\pagebreak[4]

\section{Evo-devo and transposable-elements}

\subsection{The evo-devo method}
 
Given a certain genome embedded in a single cell put on the grid, we have now a set of rules to generate a development and obtain a final shape. A naturally arising question is: how do we choose the genome in order to produce a predefined target shape? The answer is: we leverage the power of evolution through the use of a Genetic Algorithm (GA). In this case the GA evolves a population of genomes which guide the development of the shape starting from a single cell (zygote) initially present on the grid, for a number of generations.

At each generation, all genomes in the population (one at a time) guide the development of the shape from the zygote stage to the final phenotype, whose adherence to a target shape is employed as fitness measure. This operation is repeated for all genomes, so that eventually each genome is assigned a fitness value: based on this value the genomes are then selected and randomly mutated, to produce the new population. This cycle is repeated until a satisfactory level of fitness is reached.
 
The coupling of the model of cellular development and the Genetic Algorithm becomes an evo-devo method to generate 2d or 3d shapes. ``In silico'' experiments (i.e. computer simulations, see examples in \colorbox{figdr}{figures 12-13}) have proved the effectiveness of the method in ``devo-evolving'' any kind of shape, of any complexity (in terms of e.g. number of cells, number of colours, etc.). If the complexity of such shapes is interpreted as a metaphor for biological complexity, we can conclude that this method has the potential to generate the complexity typical of living beings. The power of the method essentially depends on the features of the model of development and in particular is to be reconducted to the presence of a homogeneous distribution of driver cells, which keeps the shape ``plastic'' throughout development. On the other hand, the speed of the evolutionary process is also safeguarded by a special procedure, which will be described next.  

\subsection{Germline Penetration}
 
In order to develop a given shape, the evolution implemented by the genetic algorithm has to invent developmental genes whose MOS sequence matches a MOC sequence belonging to one of the driver cells present in the shape volume. If the MOS sequence is too long, the size of the space the GA has to search causes the evolutionary process to slow down considerably (being the slow down proportional to the difficulty in guessing the right sequence). A countermeasure consists in ``suggesting'' MOS sequences to the GA likely to match existing MOC sequences. If we suggest the GA to use as genes' MOS sequences some of the MOC sequences generated during the development of the shape, the match is guaranteed.
 
This idea is implemented in a procedure called ``Germline Penetration'', executed at the end of the development of each individual shape. Germline Penetration copies at random (some) MOC sequences generated during the development of the shape onto MOS sequences of developmental genes contained in a special copy of the genome called ``germline'' genome, distinct from the genome incorporated in all cells. The germline genome, after reproduction and the application of mutations and recombinations, is destined to become the (``somatic'') genome of the individuals of the next generation, in which it will again be embedded in all cells.
 
\colorbox{figdr}{figure 14} shows an example of how Germline Penetration works. The left side of the picture shows the sequence of development for a shape belonging to a ``species'' X, at generation K. For this species development consists of a single change event, in which the zygote proliferates in step 1 and a number of new driver cells and MOC sequences are generated: B, C, D, etc. To further develop the shape, some new developmental genes have to appear in the genome, directed onto these new MOC sequences, i.e. the MOS fields in their left parts have to be equal to such MOC sequences.
 
As mentioned above, guessing the right MOS sequences is a very difficult task for the GA: Germline Penetration intervenes copying some of the new MOC sequences into MOS sequences of developmental genes present in the germline genome, genes that will appear in the genome of next generation's individuals. Evolution, whose objective is to produce individuals with a high fitness level, is now provided with ``good'' genes' left parts, i.e. left parts containing MOS sequences that are guaranteed to match existing MOC sequences. It now has to optimise the relevant right parts, a process that can take several generations.

The optimisation of the right part is completed after H generations (at generation K+H) for the individual shown on the right side of \colorbox{figdr}{figure 14}: two developmental genes now trigger MOC sequences D and E (which have been previously transplanted into the genome by Germline Penetration) to carry out as many change events in step 2: the individual belongs to a new species, Y. The new MOC sequences generated as a result of the new change events occurred in step 2 are again transferred to the germline genome to be embedded into the genome of the offspring and the whole cycle repeats itself.
 
The genes with the copied MOS sequences are initially set as inactive (their switch is set to 0), as they would otherwise all become active with a non-optimised right part, triggering random change events and causing a major disruption in development (and an abrupt decrease in the individual's fitness). Their activation, obtained through a ``flip'' of the promoter, is left to a subsequent genomic mutation: in this way, evolution has the time to produce a good right part. As a consequence, at any given time in the course of evolution of any individual, most developmental genes in the genome are inactive. These genes, in analogy with real genomes, can be defined as ``junk'' genes. 

\subsection{Transposonal theory of heredity}

Transposable elements (TE), or transposons, first discovered by B. McClintock \cite{EB50MC}, are DNA  sequences that can move around to different positions in the genome. During this process, they can cause mutations, chromosomal rearrangements and lead to an increase in genome size. Major subclasses of transposons are represented by DNA transposons, LTR retrotransposons, long interspersed nuclear elements (LINE's), and short interspersed nuclear elements (SINE's). Despite representing a large genomic fraction (30-40\% in mammals), no clear function has yet been identified for transposons, which have therefore been labelled as ``junk DNA'', until recent research suggested they could indeed have a biological role. Transposable elements have been studied along two main dimensions: evolution and development.
 
A first line of research has been concerned with the dynamics of TE diffusion in genomes across multiple generations, as can be evidenced through modern genome-wide analysis techniques. Transposable elements appear to be temporally associated with major evolutionary changes \cite{EB05FL} \cite{EB10OG} and many elements are present only in specific lineages (Alu in primates for example), suggesting a causal link between the appearance of such elements in the genome and the evolutionary change that originated the lineage. Lineages whose genomes are not subject to intermittent TE infiltrations are more static from an evolutionary point of view. A specific increased activity of some TE families in the germline has been observed \cite{EB07MM}, in agreement with the hypothesised evolutionary role of transposons. As a matter of fact, only transposons that become fixed in the germline can be passed on to the next generation and have trans-generational effects.
 
Transposable elements were initially characterised as elements controlling phenotypic characteristics during development in maize \cite{EB50MC}, when they can insert themselves near genes regulating pigment production in specific cells, inhibiting their action and making the cells unable to produce the pigment. This regulatory role is supported by further research indicating that (in human) transposons tend to be located near developmental genes \cite{EB07LB}. Recent observations corroborate the view that TEs are active in somatic cells, opening the possibility that they can bring diversity among somatic cells with the same genome \cite{EB07CL}. All together this evidence suggests that TEs can be grouped in two categories: i) TEs that are active in somatic cells during development and ii) TEs that are active in germline cells and become fixed in the genome.
 
According to Epigenetic Tracking, the cycle shown in \colorbox{figdr}{figure 14} represents the evo-devo core of multicellular life. The central role played by Germline Penetration in our model lead us to hypothesise the existence of a similar procedure also in real biological systems. The main palyers in such ``biological Germline Penetration'' would be the biological counterparts of MOC and MOS sequences (that for simplicity will still be called MOC and MOS), which we imagined implemented in nature as sets of transposable-elements. 

Following this idea, we can picture a situation in which new MOC sequences, created in proliferation events, exit the driver cells in which they are stored, reach the circulatory system and make their way into the germline genome, where they become incorporated into left parts of developmental genes, to be passed on to the individuals of the next generation. More realistically, the trasposons would insert in the genome in random positions. The genomic region falling under the regulatory domain of the inserted sequence would then come under selective pressure and become a ``biological right part'' of a biological developmental gene. After a number of generations evolution finds good solutions for such right parts and development can move ahead.
 
The mechanism described has interesting evolutionary implications. Such implications can be deducted considering that, whenever a driver cell is triggered to proliferate by a developmental gene, a ``wave'' of new driver cells, each with its own MOC sequence, is created in the body of the (new) species. The action of Germline Penetration translates this wave of new MOC sequences in somatic cells into a corresponding wave of new MOS sequences spreading in the germline genome and subsequently in the somatic genome of the next generation.
 
The moments of such events during evolution coincide with milestones in which major changes occur to the evolving species, causing new body parts or features to appear. In other words, the spreading in the genome of waves of new sets of transposable-elements in the course of evolution corresponds to moments in which new branches (new species) are generated in the ``tree of life''. This is made possible by the action of Germline Penetration, that implements a flow of genetic information from somatic cells to germline cells, to be passed on to future generations in a non-Mendelian fashion. Such predictions made with our model on purely theoretical grounds, appear to be confirmed by experimental evidence \cite{EB10OG}. 
 
One point which is worth stressing is the following. The experimental evidence reported in \cite{EB10OG} suggests that the spreading of new transposon families in the genome and the occurrence of major changes in the relevant lineage are simultaneous events, implying that the colonisation of the genome by the transposons ``drive'' the change in the lineage. The interpretation of this phenomenon provided by our model is somewhat different. In fact, the model foresees that the spread of new transposon families in the genome is an event which comes immediately (in evolutionary terms) {\em after} the change, not before.
 
More precisely, the precise sequence of events triggered by a change in the lineage is the following:
\begin{enumerate}
\item change in the lineage: creation of a new body part in a species (and hence {\em de facto} creation of a new species);
\item creation of new driver cells in that body part and of the relevant MOC sequences;
\item transfer of the MOC sequences (implemented in nature as sets of transposons) from the driver cells to the germline genome by Germline Penetration and hence generation of a wave of transposons spreading in the genome.
\end{enumerate}
If the sensitivity of the analysis techniques used to estimate the age of DNA sequences were able to confirm such somewhat counterintuitive prediction, this would represent a clear indication in favour of the proposed model. 
 
Sperm-mediated gene transfer (SMGT) \cite{EB05SS} is a procedure through which new genetic traits are introduced in animals by exploiting the ability of spermatozoa to take up exogenous DNA molecules and deliver them to oocytes at fertilisation. The reverse-transcribed molecules are propagated in tissues as low copy extrachromosomal structures, able to inducing phenotypic variations in positive tissues, and transferred from one generation to the next in a non-Mendelian fashion. Experimental evidence suggests that sperm-mediated gene transfer is a retrotransposon-mediated phenomenon. The ability of spermatozoa to take up DNA molecules and incorporate them into their genome could represent the mechanism used by nature to implement the final stage of Germline Penetration, namely the phase in which MOC sequences enter the germline genome. Epigenetic Tracking provides the biological meaning of the properties which make SMGT possible.
 
In conclusion, we have seen how, in our model, Germline Penetration represents an indispensable tool to boost the evolution of multicellular structures. This point deserves emphasis: in the case of multicellular shapes, Darwinian evolution alone {\em is not sufficient }. This has been confirmed in our experiments: if Germline Penetration is disabled, evolution halts. In this perspective development and evolution appear to be two sides of the same process (linked together by Germline Penetration): the sentence ``nothing in biology makes sense except in the light of evolution'' can be reformulated as ``nothing in multicellular biology makes sense except in the light of devo-evolution''.

\begin{figure}[p] \begin{center}
{\fboxrule=0.2mm\fboxsep=0mm\fbox{\includegraphics[height=08.50cm]{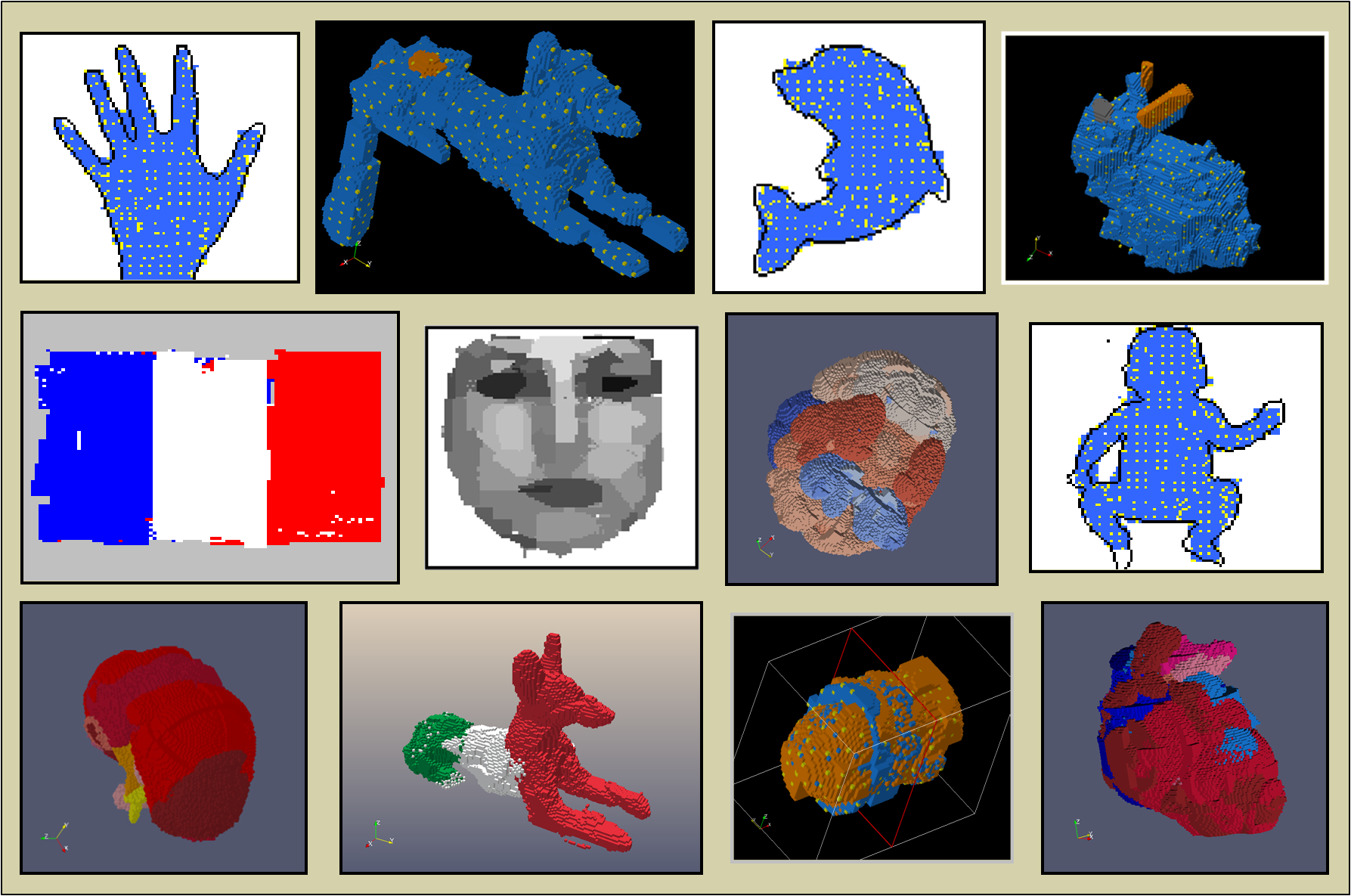}}}
\caption{Examples of phenotypes generated using Epigenetic Tracking (all figures show the final stage of the best individual evolved).}
\label{figxx}
\end{center} \end{figure}

\begin{figure}[p] \begin{center}
{\fboxrule=0.2mm\fboxsep=0mm\fbox{\includegraphics[height=9.50cm]{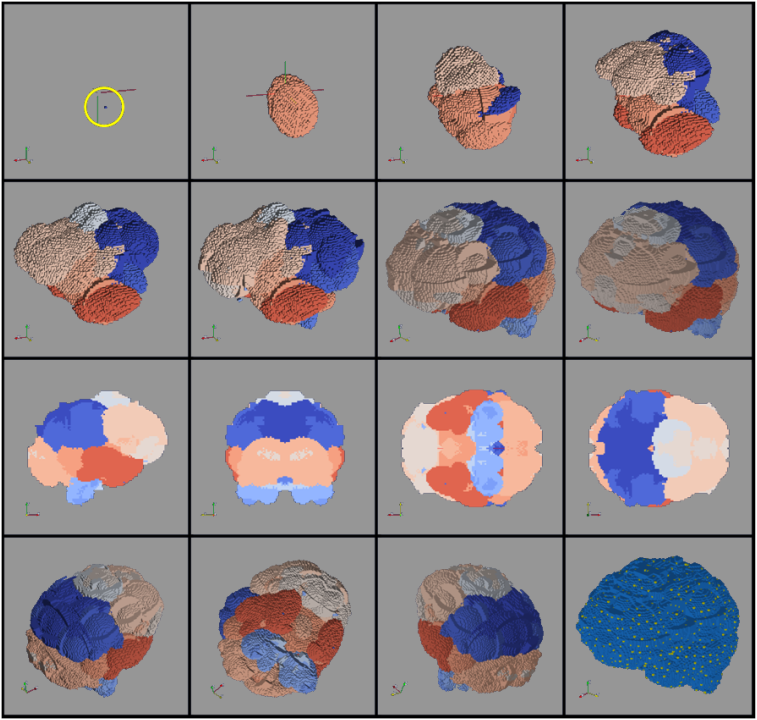}}}
\caption{Development of a 3d coloured shape representing a human brain. In the upper half the sequence of development of the best individual of the genetic population at the last generation, in the lower half some snaphshots taken from different angles.}
\label{figxx}
\end{center} \end{figure}

\begin{figure}[p] \begin{center}
{\fboxrule=0.2mm\fboxsep=0mm\fbox{\includegraphics[width=16.00cm]{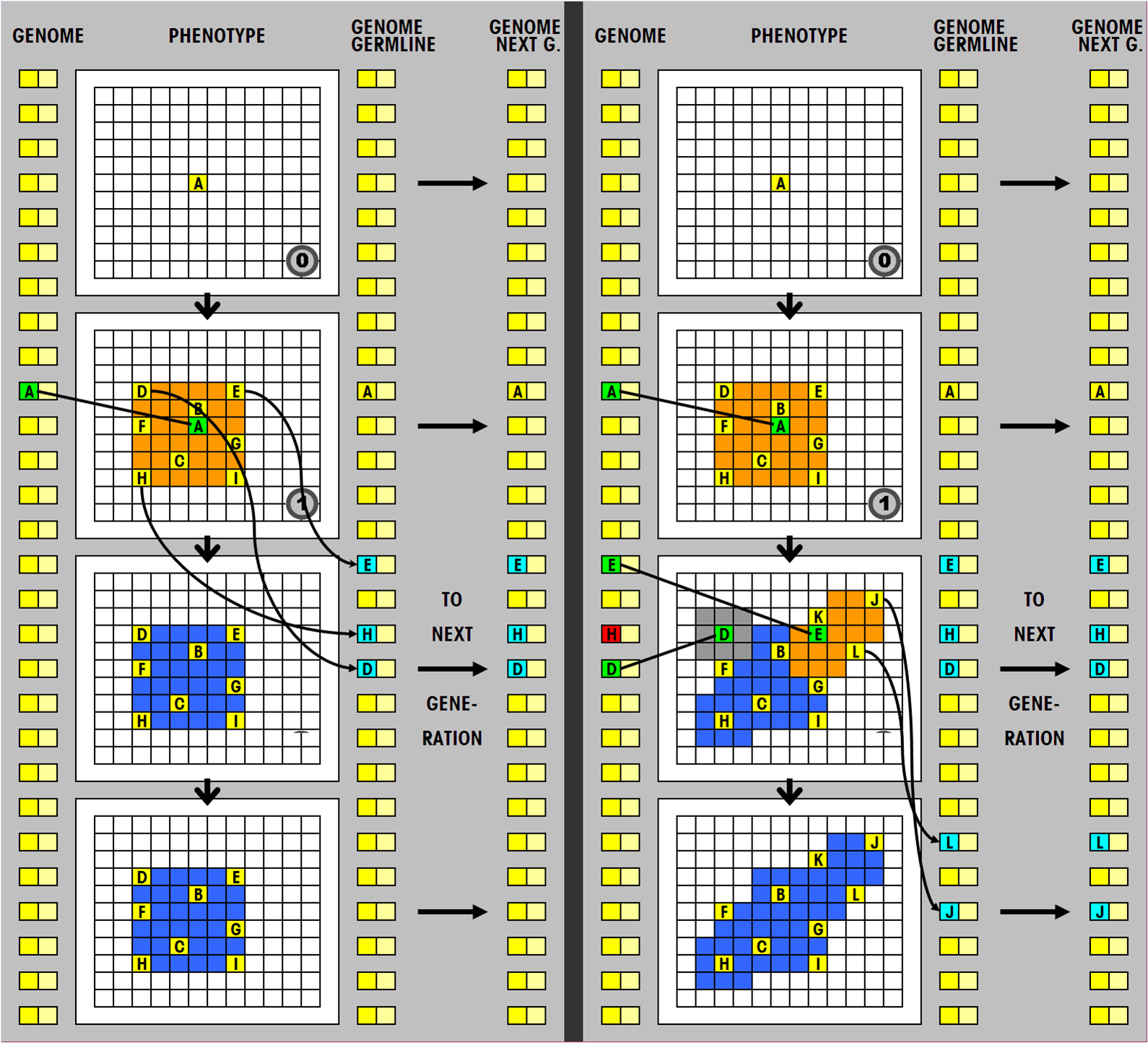}}}
\caption{Germline Penetration in action. On the left: development of a species X individual, generation K. Development stops at step 1; some of the MOC sequences generated during development leave the respective driver cells and are conveyed towards the germline genome and copied onto the MOS sequences of developmental genes. The germline genome is passed on to become the genome of the individuals of the next generation. This genome, incorporated in all cells, contains the new MOS sequences. On the right: development of a species Y individual, generation K+H. New developmental genes, derived from the transplanted elements, are present in the genome, whose MOS sequences match some of the MOC sequences generated in step 1; such genes carry on the development of species X, giving rise to a new species (Y). The new MOC sequences generated in step 2 are again copied into the germline genome and passed to the next generation.}
\label{figxx}
\end{center} \end{figure}

\clearpage

\pagebreak[4]

\section{Junk DNA and ageing}

\subsection{Facts and theories on junk DNA and ageing}

In biology the term ``junk DNA'' is used to label portions of the genome which have no function or for which no function has yet been identified. Many junk sequences appear to have been conserved over many millions of years of evolution, which seems to hint that they play an essential role: eukaryotes appear indeed to require a minimum amount of junk DNA in their genomes. In the human genome, as far as 95\% of the genome can be designated as ``junk''. Major categories of junk DNA are represented by i) introns, non-coding sequences within genes; ii) chromosomal regions composed of residues of once functional copies of genes, known as pseudogenes; iii) transposable-elements. This last category alone represents some 30-40\% of the genome of mammals.
 
A number of hypotheses have been proposed to explain the presence of junk DNA in the genome of many species. Junk DNA could represent a reservoir of material from which new genes could be selected: as such, it could be a useful tool used by evolution. Some junk sequences could be spacer material that allows enzymes to bind to functional elements. Some junk DNA can serve the purpose of preserving the integrity of the cell nucleus from a mechanical / structural viewpoint. Finally, junk DNA could be involved in regulating the expression of protein-coding genes. What is clear is that, while the amount of coding DNA appears to be similar across a wide range of species, the amount of junk DNA displays a much broader range of variation and seems indeed to be correlated with organismal complexity \cite{EC02GX}.    
 
Ageing is a process that occurs in the lifespan of most living beings. It involves the accumulation of changes in the organism over time, leading to a progressive deterioration of bodily functions. Ageing does not affect all species: in some simple species, its effects are negligible or undetectable. Moreover, the rate of the process is very different among affected species: a mouse is old at three years of age, a human at eighty years of age. Differences exist also between genders (women have on average a longer lifespan compared to men) and among single individuals. Both genetic and environmental factors seem to be involved in the ageing phenomenon. Many genes have been identified, that increase the lifespan in some species; a measure that was proved to be effective in reducing the rate of ageing across many species is caloric restriction.
 
Many theories of ageing exist. {\bf Stochastic theories} blame environmental factors that induce damage on living organisms as the cause of ageing. Within this category, the ``wear-and-tear'' theory argues that ageing is the result of damage accumulating over time (similarly to the ``ageing'' of a mechanical device). The ``free-radical theory'' is based on the idea that free radicals induce damages in cells, which at the organismal level produce the ageing phenotype. According to {\bf programmed theories} ageing is regulated by biological clocks, which trigger changes to maintenance and repair systems. Within this category, the ``ageing-clock theory'' maintains that ageing results from a programmed sequence, triggered by a clock built into the functioning of the nervous and / or endocrine systems. The shortening of the telomeres at each cell division could represent a possible physical implementation of the clock.
 
{\bf Evolutionary theories} \cite{EC02GG} consider evolutionary phenomena as the main cause for the differences in the ageing rates observed across different species. The three major evolutionary theories are the ``mutation accumulation theory'', the ``antagonistic pleiotropy theory'' and the ``disposable soma theory''. The first one states that gene mutations affecting individuals at a young age (before reproduction) are strongly selected against, while mutations to genes displaying their effect at an old age experience no selection, because the individuals have already passed their genes to the offspring. The second theory argues that strategies leading to a higher reproductive fitness and a shorter lifespan would be favoured by natural selection. The disposable soma theory argues that mutations which save energy for reproduction (positive effect), by disabling molecular proofreading devices in somatic cells (negative effect), would be favoured be natural selection. Ageing would be the cumulative result of the negative effects on the organism.

\subsection{Interpretation of junk DNA by Epigenetic Tracking}
 
In the Epigenetic Tracking framework, for any individual the set of MOC sequences (SMS) generated during development can be divided into i) MOC sequences that activate a developmental gene during development and ii) MOC sequences that do not activate any developmental gene during development. Likewise, the set of developmental genes (SDG) can be divided into i) developmental genes that are activated during development and ii) developmental genes that are not activated during development. By analogy with real cells, elements in the two categories labelled with ii) (inactive elements) can be defined as ``junk'' MOC sequences and ``junk'' developmental genes. Typically, sets labelled with ii) are much larger than sets indicated with i): in our experiments the average ratio MOC sequences used / generated is 10\% (all MOC sequences present in the organism at the end of its development, stored inside all ``yellow cells'' in the final shape, have not been used during development).
 
As it will become clear, the Epigenetic Tracking machine cannot do its job without generating a lot of junk MOC sequences. In fact, the only way to reduce the amount of unused MOC sequences consists in decreasing the ratio between driver and normal cells in proliferation events. This would cause the evolved shapes to have, at any developmental step, a sparser distribution of driver cells and, as a result, the ``sculpting tool'' would become less precise and the evolution of development would become harder. So, there seems to be a trade-off between the precision of sculpture and the density of driver cells (and hence the percentage of junk MOC sequences). Since the effectiveness in evolving shapes is Epigenetic Tracking's top priority, the precision of sculpting must be preserved: therefore, a certain amount of junk in the set of MOC sequences must be reckoned with.
 
This fact, in combination with the need to use Germline Penetration (which is an indispensable evo-devo tool and cannot be eliminated without impairing the evolutionary process), explains the presence of junk in the genome. As a matter of fact, Germline Penetration acts like a shuttle, transferring junk MOC sequences from the set of MOC sequences onto a corresponding number of junk MOS sequences in the genome. The developmental genes in which the transplant has taken place are junk genes because, as we said, their switches are initially set to 0. Without this measure, they would all become active at once with a non-optmised right part, causing disruptions in the development of the individual. In conclusion, the presence of junk information in both the set of MOC sequences and the genome is a fact which is inescapably connected to the core of the Epigenetic Tracking machine, a requirement essential to its {\em evolvability}. 
  
Recalling how MOC / MOS sequences are hypothesised to be implemented in nature as sets of transposable-elements, this phenomenon observed in our computer simulations provides an explanation for the presence of this category of junk DNA which, as we have seen, accounts for up to 40\% of the genomic content in mammals. According to this view, in the genome of a generic cell in the course of an individual's life, some MOS sequences have already been bound to corresponding MOC sequences, some will be bound in future steps and some will never be bound and can be considered true junk. We will now see how junk DNA is linked to the phenomenon of ageing. 

\subsection{Interpretation of ageing by Epigenetic Tracking}

As we know, for a given individual development unfolds in N developmental steps; at the end of it the individual's fitness is evaluated and right afterwards the genome content is handed over to the subsequent generation. The moment of fitness evaluation, that in nature can be thought to roughly correspond to the moment of reproduction, has always coincided in our experiments with the end of the simulation. On the other hand, we can imagine to let the global clock tick on and see what happens in the period after the moment of fitness evaluation. The distinction between the periods before and after fitness evaluation can be thought to correspond to the biological periods of development (say, until 25 years of age in humans -the average age of reproduction) and ageing (from 25 years of age onwards). 
 
The stock of junk elements represents a reservoir of change events that can potentially be triggered in the ageing period, after the moment of fitness evaluation, when they are by definition not affecting the fitness value. As a result, they will tend to have a random nature and their effects on the health of the individual are more likely to be detrimental than beneficial. Here, ``fitness'' has to be interpreted as the {\em reproductive} fitness of the individual (i.e. the chance that its genes will survive in the future genetic pool of the population), while ``health'' represents its ``physical'' condition. Therefore, in the ageing period, the individual's health will tend to progressively decrease over time under the action of such random events: this, we think, is the deep nature of the ageing phenomenon.
 
\colorbox{figdr}{figure 15} shows how the situation looks like at the end on an individual's development. Many driver cells are present in the body of the individual, which have not been activated during development and which can become active during the ageing period. The numbers in the circles beside each driver cell represent the values of the timers of the developmental genes which are going to be activated in that driver cell. \colorbox{figdr}{figure 16} gives an example of an ageing-related event: the driver cell bearing MOC sequence A is induced to proliferate by a developmental gene at step 64 (beyond fitness evaluation); the same developmental gene could activate / deactivate some metabolic genes, leading to a change in metabolism. The nature of these events is such that, after a few steps, the MOC sequences generated do not trigger further events and the proliferation halts.
 
Therefore (see \colorbox{figdr}{figure 17}) the set of all MOC sequences generated during an individual's development can be divided into:
\begin{enumerate}
\item The set of MOC sequences active during development (SMS-D)
\item The set of MOC sequences active during ageing (SMS-A)
\item The set of MOC sequences never active (SMS-J)
\end{enumerate}
Analogous distinction can be done also for the set of developmental genes:
\begin{enumerate}
\item The set of developmental genes active during development (SDG-D)
\item The set of developmental genes active during ageing (SDG-A)
\item The set of developmental genes never active (SDG-J)
\end{enumerate}
In this framework, ageing can be seen as a {\em continuation of development}, driven by developmental genes activated in specific driver cells after fitness evaluation.  
 
An example of ageing is reported in the upper part of \colorbox{figdr}{figure 18} for a picture representing a face: the left part shows steps 1-6, belonging to the period of development: the shape grows from the single cell stage to the mature phenotype in step 6, when the fitness is evaluated. The sequence on the right refers to the period of ageing (steps 7-12), characterised by the accumulation of random events, whose global effect causes a progressive deterioration of the quality (the ``health'') of the picture. 
 
The hypothesis according to which the evolutionary pressure acting on a developmental gene is zero if the gene becomes active after reproduction, is rather simplistic. As a matter of fact, in nature, an individual's reproductive fitness depends also on events manifesting themselves after reproduction, as also these affect the survival chances of the progeny. In other words the effect of an event on the fitness (and hence the evolutionary pressure acting upon the corresponding gene) tends to decrease as the age of its appearance -determined by the gene timer value- increases, rather than vanishing right after reproduction.
 
The lower part of \colorbox{figdr}{figure 18} shows the curve of the evolutionary pressure as a function of the timer value of genes. The pressure is high until the age of 25 years, when reproduction occurs; then it starts declining with a mild slope, since the individual has to be in good shape to look after its children, who are unable to survive alone; at the age of 50 years, the slope of decline becomes steeper, as its parental role is less important and the ex-children have children themselves. Until the age of 75 years, the individual can play a role as a grandparent; after the age of 75 years, whatever happens to our individual has no influence on the survival of its progeny: therefore, genes activated after this moment are subject to no evolutionary pressure.
 
The view of the ageing process as the progressive accumulation of random events having little or no effect on an individual's reproductive fitness provides a new interpretation for genetic diseases typically associated to the old age. Examples of such diseases are Alzheimer disease, type II diabetes, heart problems, and in general all diseases that seem to have an idiopathic or intrinsic origin (i.e. they are not caused by external agents such as viruses or bacteria) but whose onset in humans typically occurs from the 5th decade of life onwards. These diseases seem indeed to be caused by the malfunctioning of specific genes and could therefore be labelled as genetic diseases, even though they are not present at birth but manifest themselves only later in life.
 
The hypothesis we are proposing here is that the difference between the phenotipic manifestations of such diseases and the effects of ``normal'', ``healthy'' ageing are rather quantitative than qualitative. Both normal ageing and ageing-associated diseases are indeed driven by change events caused by developmental genes whose timers are set to go off in the old age: the phenotipic manifestations  related to normal ageing are only milder, more benign than those associated to the diseases. This interpretation provides a straightforward explanation on why the temporal patterns of ageing and these diseases are coincident: they are basically the same thing. 
 
As we mentioned, a slow-down in the pace of ageing can be obtained through caloric restriction. One possible explanation of this experimental evidence is that caloric restriction directly affects the functioning of the global clock, which becomes slower. As a result, all developmental genes set to go off in the ageing period are delayed, all by the same amount. Some genes, which are known to slow-down ageing in some species, could exert an influence on the cellular molecular machinery which transduces the clock (likely implemented as a protein) into the nucleus. In this way the value of the clock perceived inside the cell will be lower.       
 
In summary we can say that the theory proposed belongs to the class of evolutionary theories of ageing and is basically consistent with the mutation accumulation theory. The novelty is represented by two elements:
\begin{enumerate}
\item this theory sees ageing as the continuation of development, thus bringing together development and ageing under a unified theoretical framework, based on the functioning of driver cells;
\item this theory specifies in a very precise and ``operational'' way the cellular bases of the ageing phenomenon, linking the behaviour of driver cells and developmental genes to the physical signs of ageing.
\end{enumerate}
To our knowledge, the simulation reported in \colorbox{figdr}{figure 18} is one of the very few present in the literature, made possible by the fact that the underlying theory is based on a computer model susceptible of being simulated.
 
We wish to conclude this section dedicating a final comment to the role played by elements inactive during development. We have shown how a part of these elements (SMS-A and SDG-A) is actually devoted to cause random events in the ageing period, relegating to SMS-J and SDG-J the role of true junk. On the other hand, elements are free to move between sets: the sets of inactive elements represent therefore a reservoir of potential new events occurring during development and a tool to explore new evolutionary paths. These considerations bring us to deducting a direct link between the evolvability of a species and its susceptibility to ageing, being both aspects mediated by the presence of a big stock of junk. The fact that bats have unusually small genomes (i.e. little junk) and display a remarkably long lifespan (i.e. they appear to age less) among mammals of comparable dimension \cite{EC95BL}, appears to be consistent with this hypothesis.  

\begin{figure}[p] \begin{center}
{\fboxrule=0.2mm\fboxsep=0mm\fbox{\includegraphics[height=06.00cm]{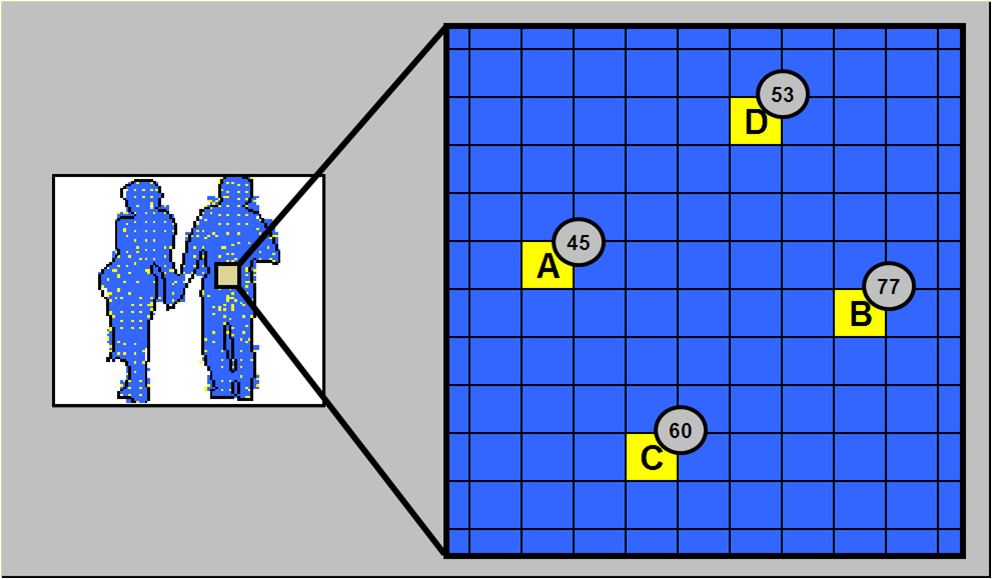}}}
\caption{Ageing driver cells at the end of development. The timer value of the developmental gene which is going to be activated is indicated in the circle.}
\label{figxx}
\end{center} \end{figure}

\begin{figure}[p] \begin{center}
{\fboxrule=0.2mm\fboxsep=0mm\fbox{\includegraphics[width=13.50cm]{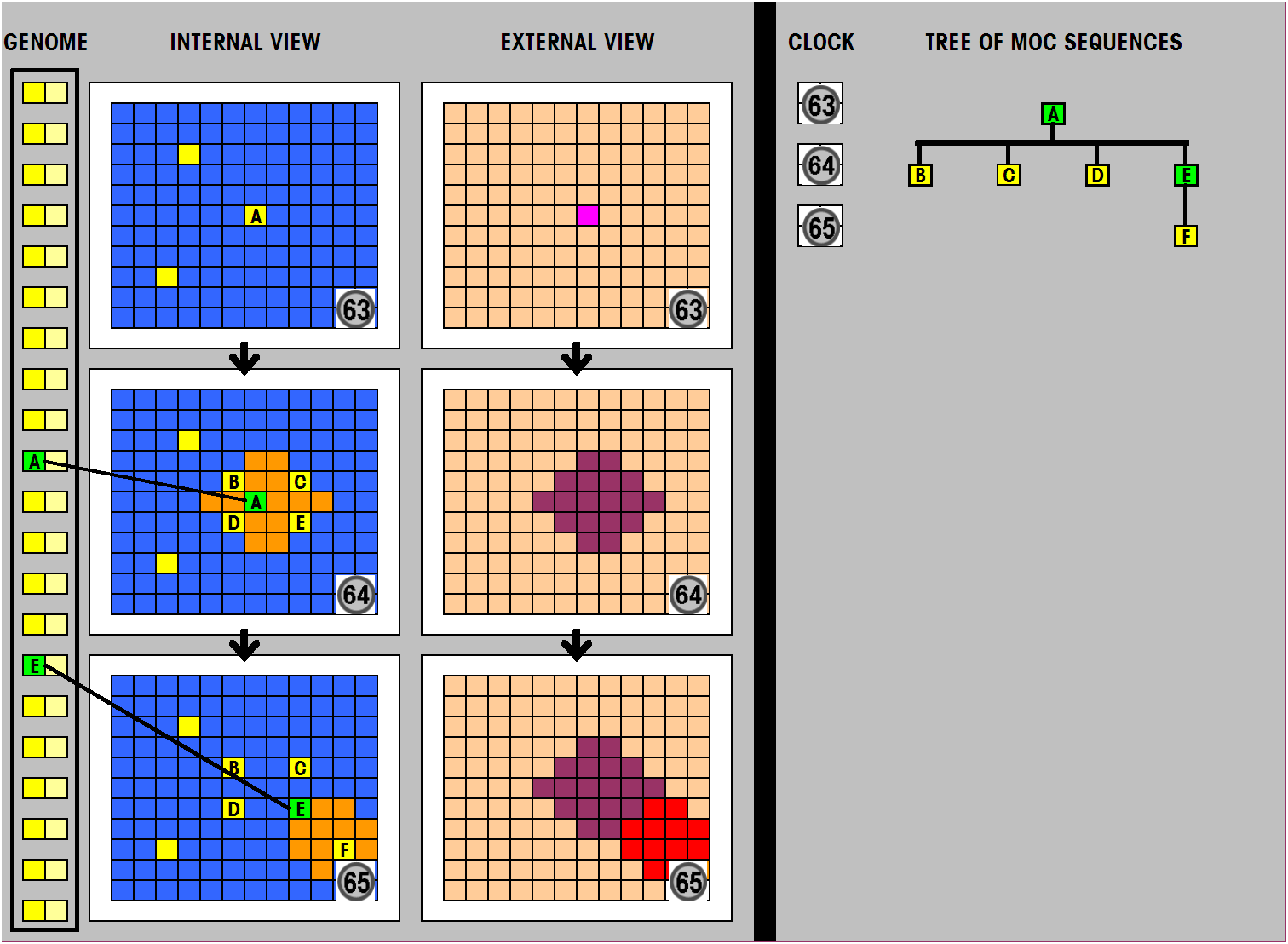}}}
\caption{Example of an ageing-related proliferation in a driver cell, induced to proliferate by a developmental gene during the ageing period. The same developmental gene could activate and de-activate the promoters of some metabolic genes, leading to a change in metabolism. After step 65 the MOC sequences generated do not trigger further events and the proliferation halts. The effects contribute to the ageing phenotype.}
\label{figxx}
\end{center} \end{figure}

\begin{figure}[p] \begin{center}
{\fboxrule=0.2mm\fboxsep=0mm\fbox{\includegraphics[height=05.00cm]{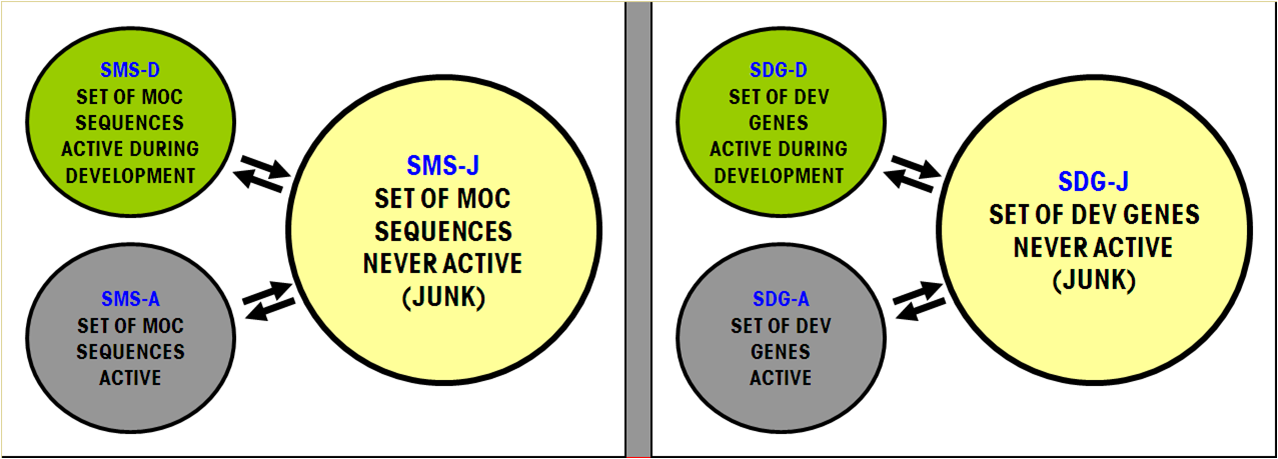}}}
\caption{On the left: sets of MOC sequences which activate a gene during development (SMS-D), activate a gene during ageing (SMS-A), never activate any gene (SMS-J). On the right: sets of developmental genes which are activated during development (SDG-D), are activated during ageing (SDG-A), are never activated (SDG-J).}
\label{figxx}
\end{center} \end{figure}

\begin{figure}[p] \begin{center}
{\fboxrule=0.2mm\fboxsep=0mm\fbox{\includegraphics[width=15.00cm]{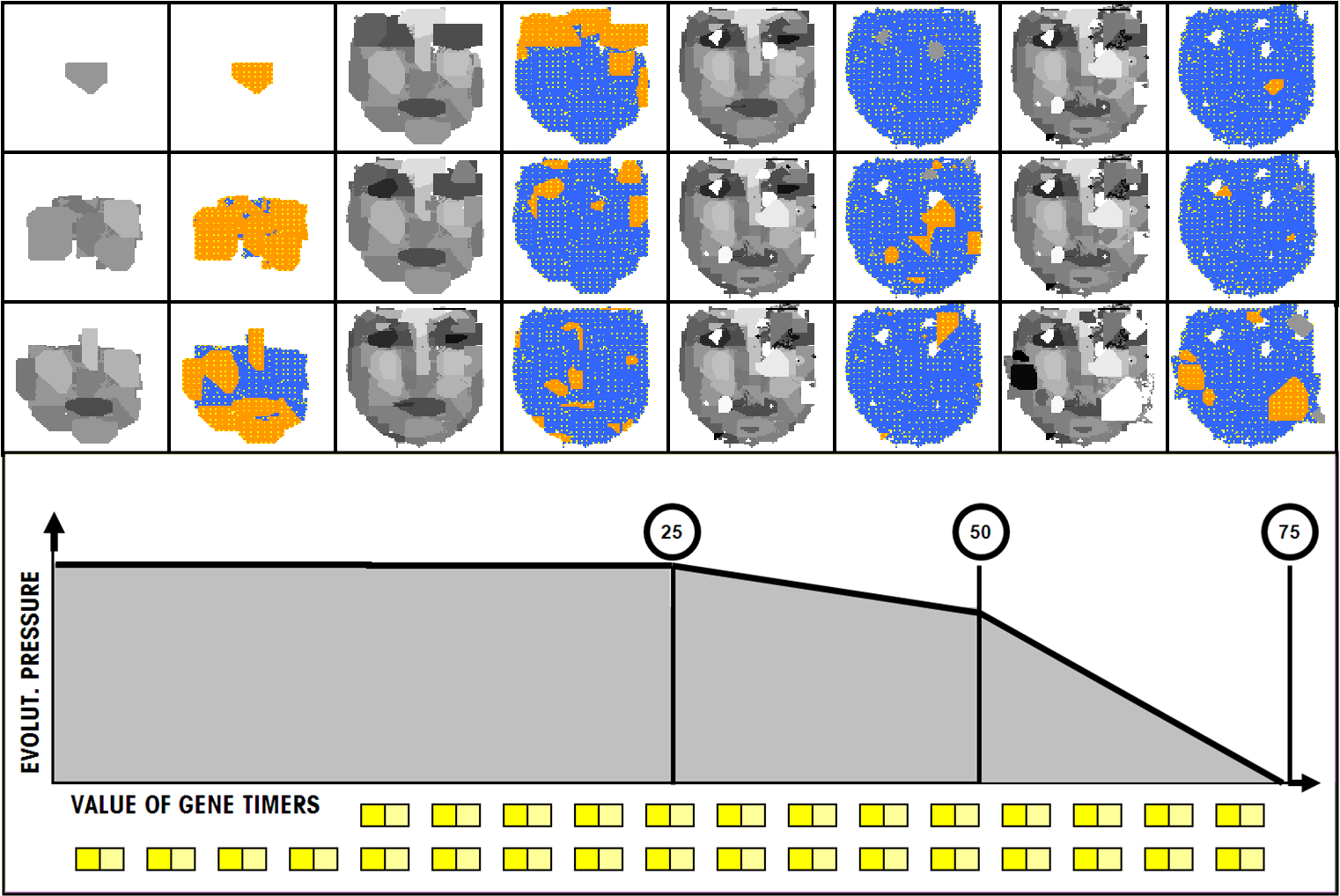}}}
\caption{In the upper part: example of ageing for a ``face''. On the left the period of development (steps 1-6): the shape grows from a single cell to the mature phenotype in step 6, when fitness is evaluated; on the right the period of ageing (steps 7-12): the picture quality deteriorates steadily under the action of random developmental genes. In the lower part: the evolutionary pressure acting upon a development gene is an inverse function of the value of the gene's timer. Genes with timer values (referred to human) lower than 25 years are subject to high evolutionary pressure, genes with timer values greater than 25 years are subject a steadily decreasing pressure, until their effects become random.}
\label{figxx}
\end{center} \end{figure}

\clearpage

\pagebreak[4]

\section{Cancer}

\subsection{Teratomas}

In this section we will analyse a possible malfunction of the model and we will show how such a malfunction gives origin to a phenomenon that mimics a particular kind of tumour called a teratoma. A teratoma is a tumour with tissue or organ components resembling normal derivatives of all three germ layers. The tissues of a teratoma, although normal in themselves, may be quite different from surrounding tissues, and may be highly inappropriate, even grotesque: teratomas have been reported to contain hair, teeth, bone and very rarely more complex organs; usually, however, a teratoma does not contain organs but rather tissues normally found in organs such as the brain, liver, and lung. Teratomas are thought to be present at birth, but small ones often are only discovered much later in life.
 
In our model, a certain body part of an organism can be generated by a single driver cell that, once activated, proliferates, generating other driver cells; some of these driver cells get in turn activated, proliferating and generating other driver cells etc. (the same holds true for the entire organism). This process, which has been shaped by evolution to occur in a precisely orchestrated fashion, presupposes that each driver cell, at the moment of activation, finds itself in the right position, surrounded by the right cellular micro-environment: only in this case is the cascade of events originated from the driver cell's activation capable, along with physics, of generating the relevant body part.
 
This delicate mechanism can be perturbed in many ways. We will now focus our attention on a case characterised by a mutation that, at step GC(J), turns the MOC sequence (J) of a certain driver cell C(J), positioned at point P(J), into another MOC sequence (K); if the MOC sequence K is not generated during normal development, or if it is generated but never activated, nothing happens. If, on the contrary, the MOC sequence K does become active during normal development to produce a certain body part -say at step GC(K), when cell C(K) finds itself at point P(K)- as a result of the mutation the cascade of events destined to give rise to such body part will start from both point P(K) at step GC(K) -right place and moment- and point P(J) at step GC(J) -ectopic place, wrong moment.
 
Being activated in the wrong place and moment, cell C(J) is not surrounded by the right micro-environment: as a result, the cascade of events originating from C(J) will only manage to mimic the development of the relevant body part in a grotesque fashion. \colorbox{figdr}{figure 20} shows a simulation of a teratoma, occurring to a human embryo, whose normal development is shown in \colorbox{figdr}{figure 19}. At step 4, the MOC sequence (J) of the driver cell marked with the circle (C(J)) is mutated into the MOC sequence of the zygote (K) (hence GC(J)=4 and GC(K)=1); as a result, the development of the whole embryo starts over again from cell C(J): the cell proliferates, generating other MOC sequences some of which, as in normal development, trigger other proliferation events and so on.
 
But, since in this case the zygote and all other MOC sequences originated from it are in ectopic positions and are surrounded by wrong environments, while the different cell types (represented by different colours) continue to be created, the interactions with other cells -mediated by physics- prevent them from being arranged in the correct patterns; instead, an amorphous mass of differentiated cells is produced. The kind of mutation reported in this simulation is only one among endless possibilities; another possible path leading to a teratoma is the following: the MOC sequence belonging to a driver cell of the developing liver is turned into the MOC sequence of the driver cell which in normal development is the precursor of the hand; as a result, the mutated driver cell will try to generate the hand etc. We will now turn our attention to carcinogenesis in the general case.

\subsection{Facts and theories on carcinogenesis}
 
Carcinogenesis is the process by which normal cells are transformed into cancer cells. Bases on studies of skin cancer, the process of carcinogenesis is traditionally divided into three phases: initiation (linked to chemicals or physical insults that induce permanent alterations to DNA), promotion (the proliferation of the initiated cell induced by subsequent stimuli) and progression (the stepwise transformation of a benign tumour into a malignant one). The ``standard theory'' states that carcinogenesis is a multi-step process that can take place in any cell, driven by damage to genes that normally regulate cell proliferation. This in turn upsets the normal balance between cell proliferation and cell death and results in uncontrolled cell division and tumour formation. The standard theory is referred to as the ``multi-hit'' hypothesis; a critical point inherent to it is that not all cells of a tumour seem able to induce a secondary tumour when injected into nude mice \cite{ED08VL}.
 
This latter piece of evidence is addressed by a more recent theory which traces back the origin, the maintenance and the spread of a tumour to a relatively small subpopulation of cells called {\bf cancer stem cells (CSCs)}, whereas the bulk of the tumour would actually be composed of non-tumorigenic cells that, deprived of the cancer stem cells, would quickly shrink and disappear. CSCs possess characteristics associated with normal stem cells, specifically the ability to give rise to all cell types found in a particular cancer sample; CSCs may generate tumours through the stem cell processes of self-renewal and differentiation into multiple cell types. The implications of this hypothesis for therapy cannot be overstated: conventional chemotherapies kill differentiated or differentiating cells, which form the bulk of the tumour but are unable to generate new cells; a population of CSCs, which gave rise to it, could remain untouched and cause a relapse of the disease.
 
Since cancer is a disease of genes, the attempt to link cancer or specific cancers to patterns of gene mutations, consistently found in all tumour samples, would seem logical and well-founded. A few cancer-related genes, such as p53, do seem to be mutated in the majority of tumours, but many other cancer genes are changed in only a small fraction of cancer types, a minority of patients, or a subset of cells within a tumour. Although the effort to reconduct tumour formation to subsets of mutated genes in 100\% of cases has so far been unsuccessful, it is nonetheless undeniable that {\em correlations} between different tumours and specific patterns of mutations exist. Individual genes are mutated in percentages that are tumour-specific, e.g. the rb gene is mutated in 50\% of colorectal cancers, in 30\% of adenocarcinomas, etc.: these correlations represent evidence a theory of carcinogenesis has to explain.
 
Aneuploidy, defined as the occurrence of one or more extra or missing chromosomes leading to an unbalanced chromosome complement, is another characteristic feature of most tumours. ``If you look at most solid tumours in adults, it looks like someone set off a bomb in the nucleus. In most cells, there are big pieces of chromosomes hooked together and duplications or losses of whole chromosomes.'' \cite{ED02HW} \cite{ED04GX} Aneuploidy has often been regarded as an alternative way to induce loss of heterozygosity in key cancer-related genes and to render them non-functional. Nevertheless, the phenomenon is so widespread that it appears indeed to be an intrinsic characteristic of cancer cells, linked to the core of the carcinogenic process in a more profound, yet unknown, way.   
 
The presence in tumours of cells of different types and / or having different degrees of differentiation is a well documented phenomenon, coherent with the cancer stem cell theory. There is increasing evidence that diverse solid tumours are organised in a hierarchical fashion and their growth is sustained by a distinct subpopulation of CSCs. This fact is difficult to explain by the standard theory, which postulates that all tumour cells are derived clonally from a single cell, the first to have accumulated the number of hits required to achieve the malignant transformation. According to the standard theory, genotypic and phenotypic diversity within a tumour can only be realised through the effect of subsequent mutations. Evidence, on the other hand, points to the existence of a structured hierarchy of cell types, displaying variable levels of separation from their healthy counterparts.   
 
Cancer is a disease of the old age. The temporal patterns of ageing and cancer appear indeed to be perfectly superimposed: cancer is a rare occurrence in the young and becomes more and more common with age progression. The most common explanation of this phenomenon is based on the ``multi-hit'' hypothesis, according to which multiple ``hits'' to DNA are necessary to cause cancer. In this perspective, the chance that a cell accumulates the number of hits required for transformation increases with the age of the individual: thus, the late onset of cancer simply reflects the time necessary for a series of rare events to occur. The fact that a 6th power law fits well statistical data on cancer prevalence and age, seems to suggest that six independent hits are on average required for carcinogenesis.

\subsection{Interpretation of carcinogenesis by Epigenetic Tracking}

As we said, at the end of an individual's development many junk driver cells are present, as well as many junk developmental genes: this stock of junk represents a reservoir of proliferation events that can potentially be triggered in the ageing period. Since the effects of these events on the survival chances of the individual's progeny are weak and steadily decreasing, they will tend to have a random nature and their impact on the individual's health are more likely to be detrimental than beneficial. Each event has a non self-sustaining nature and hence a limited duration: overall, their cumulative effect translates to a slow, ``benign'' decrease of the individual's health, that we call ageing.
 
Now, the stage for a dangerous scenario is set if a fault arises in one of such ``ageing'' driver cells, affecting the mechanism used to generate new MOC sequences during proliferation events. As we have seen, a proliferating driver cell (the ``mother'' driver cell) first produces new normal cells: some of these will afterwards turn themselves into driver cells (the ``daughter'' driver cells) in response to chemical signals received from the surrounding driver cells (\colorbox{figdr}{figures 6 and 7}). The mechanism for driver transformation is inherited from the mother driver cell: if it is faulty, the fault will be inherited by the newly created normal cells, affecting their capacity to generate new MOC sequences. 
 
This mechanism can be damaged in many ways: in one possible variant the damage can be such that one or more daughter driver cells end up having the same MOC sequence of the mother (\colorbox{figdr}{figure 21}). Since the activation of a specific developmental gene in a driver cell depends on the MOC sequence, in this situation the mother cell and its daughters with identical MOC sequence are forced to trigger the activation of the same developmental gene. Since this gene codes for proliferation, the result is a chain reaction of proliferation events, the mark of carcingenesis.
 
As we have seen, the MOC-generation mechanism is realised through the transduction of the chemical signals released by the surrounding driver cells (see \colorbox{figdr}{figure 22}) into the cellular compartment in which the MOC is stored. Here they are processed by a dedicated piece of cellular machinery (the red box), which generates a new sub-sequence appended to the existing MOC sequence, to produce the new MOC sequence for the new driver cell. The hypothesis we wish to put forward here is that this ``red box'' is implemented in nature by gene regulatory networks in which a key role is played by tumour-suppressor (TS) genes. The new MOC sequences, in other words, are determined by the interplay of the products of TS genes.
 
In the biological interpretation of Epigenetic Tracking, the set of MOC sequences necessary to induce differentiation are tissue-specific: the set of MOC sequences, for instance, that induce differentiation in skin progenitor cells is different from the set of MOC sequences needed to induce differentiation in, say, gut progenitor cells. The same must be true for the sets of TS genes involved in the generation of those MOC sequences. Given, for instance, six TS genes, (TS1, TS2, TS3, TS4, TS5, TS6), a subset of the genes, say (TS1, TS2, TS5), could be necessary to generate the MOC sequences needed for skin differentiation, while a different subset, say (TS4, TS5, TS6), could be necessary to generate the MOC sequences needed for gut differentiation. If a sufficient number of TS genes are rendered non-functional (through mutations), the MOC-generation mechanism, responsible for differentiation, is damaged. This situation is hypothesised to correspond to initiation, the first stage of carcinogenesis.  
 
The actual proliferation of the driver cell, which corresponds to the biological stage of promotion, occurs once all conditions required for gene activation are met. In particular, the activation of the relevant developmental gene is postponed until the clock reaches the timer value. It is important to note that, if the cell does not proliferate, the effects of the damage to the MOC generation mechanism do not manifest themselves, even if present. This can lead to a long latency period, in which the potential for carcinogenic proliferation is present but not elicited. Finally, progression, the third and last phase of carcinogenesis, is hypothesised to be driven by further mutations occurring to oncogenes, which confer additional powers to the already transformed cells, e.g. the capacity to infiltrate tissues and to produce distant metastases.
 
Since the set of MOC sequences necessary to induce differentiation is tissue-specific and MOC sequences are generated by TS genes, it is not surprising to find that the set of mutated TS genes is cancer-specific. A fully-working differentiation machinery requires all TS genes to be functioning. With reference to the example above (where TS1, TS2 and TS5 are necessary for skin differentiation), if, say, TS1 gets damaged, the daughter driver cells become less different from their mother; if also TS2 gets damaged, the degree of differentiation is further reduced. It is important to point out that, while full differentiation occurs in a unique way (all involved TS genes have to be functional), a reduced form of differentiation can be realised in many different ways, corresponding to all possible different subsets of non-functional TS genes. 

The biological interpretation of MOC sequences as sets of transposable-elements offers an easy explanation of the ``bomb in the nucleus'' effect. In a healthy driver cell, as we have seen, a MOC sequence jumps into a precise position in the genome, activating a specific developmental gene. A natural MOC sequence, as we said, is as a set of DNA elements containing transposons, that jump into precise locations on chromosomes, thereby activating specific genes. We can imagine how corrupted transposons would jump into wrong chromosomic loci, where the chromatin structure is not suited to host them. As a result, the chromosome would break off, causing the typical scenario observed in most cancerous cells, the ``bomb in the nucleus'' effect. This would in turn contribute to accelerate the mutagenic process, disabling further TS genes and leading to even worse scenarios. 
 
The presence in tumours of cells having different degrees of differentiation and/or of different cell types is another fact which is easily accounted for by our model. As a matter of fact, along with driver cells bearing the same MOC sequence of the mother, the key mark of carcinogenesis, other driver cells with a normal MOC sequence may be present. This because the subset of TS genes necessary for the generation of their MOC sequences has not been affected by mutations. An example of this is shown in \colorbox{figdr}{figure 23} where, along with the purple cells, also cells of a different type (in this case the red cells) are present, leading to a heterogeneous mix of cell types. It is quite natural to think that the grade and the rate of growth of a tumour are linked to the share of driver cells having the same MOC sequence of the mother in the hierarchy.    
 
The proposed theory provides also a quite straightforward explanation for the relation between cancer prevalence and age. Indeed, the very same events triggered in the ageing period are hypothesised to contribute to the ageing phenomenon (if the MOC generation machinery is intact) or give rise to a tumour (if the MOC generation machinery is damaged). As a result, the temporal patterns of ageing and cancer are coincident. The theory sheds light also on the phenomenon of latency, i.e. the fact that the exposure to mutagenic substances (e.g. tobacco smoke) and the appearance of the related tumour (e.g. lung cancer) can be separated by many years. As a matter of fact, even if the damage to the driver cell's MOC generation mechanism occurs early in life, for its effects to become manifest we need to wait until a proliferation event is triggered on the relevant cell by a developmental gene: if that gene's timer is set to 60 years of age, the tumour will not appear until that moment. 

\begin{figure}[p] \begin{center}
{\fboxrule=0.2mm\fboxsep=0mm\fbox{\includegraphics[height=09.00cm]{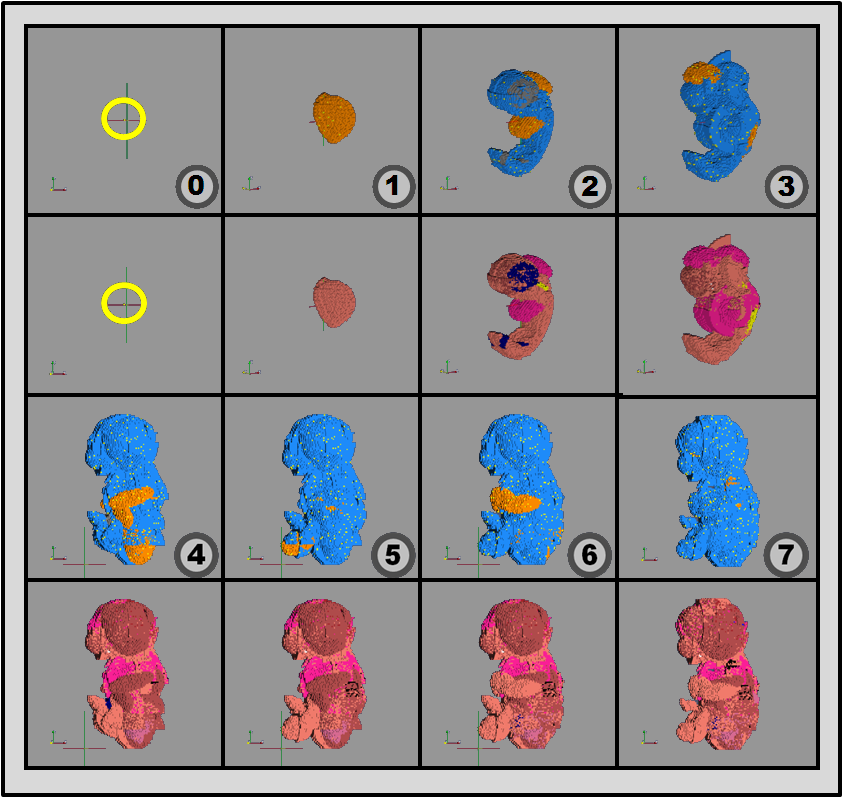}}}
\caption{Simulation of the development of a human embryo from a single cell.}
\label{figxx}
\end{center} \end{figure}

\begin{figure}[p] \begin{center}
{\fboxrule=0.2mm\fboxsep=0mm\fbox{\includegraphics[height=09.00cm]{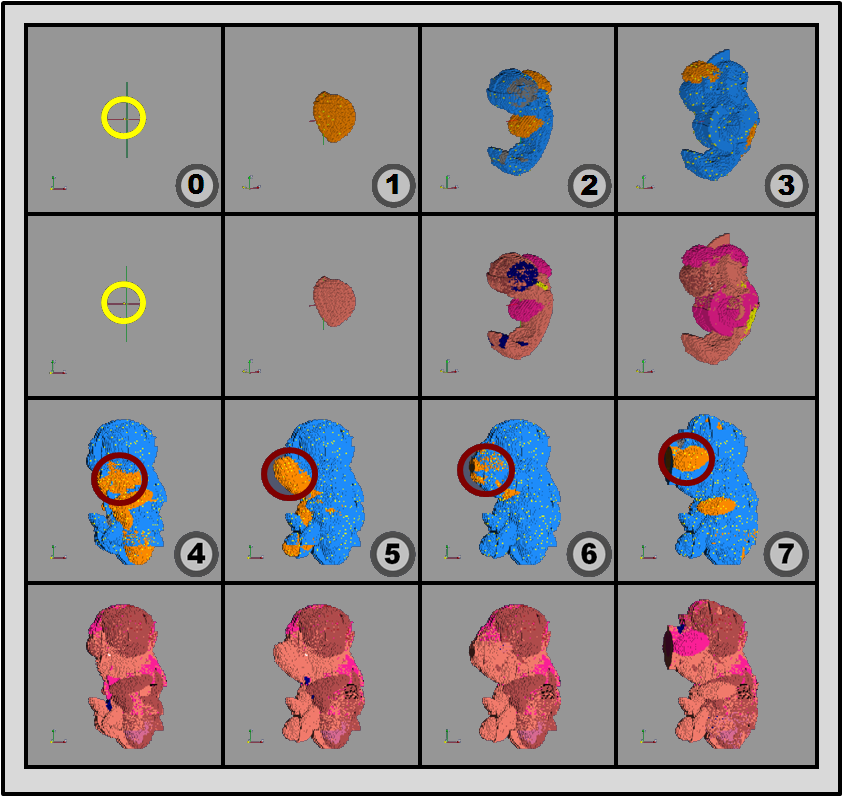}}}
\caption{Simulation of a teratoma on the embryo of \colorbox{figdr}{figure 19}. At step 4 the MOC sequence belonging to the driver cell circled in red is turned into the MOC sequence of the zygote: as a consequence the development of the whole embryo starts over from the point indicated, producing a shapeless mass of cells in the neck region, composed of differentiated cells.}
\label{figxx}
\end{center} \end{figure}

\begin{figure}[p] \begin{center}
{\fboxrule=0.2mm\fboxsep=0mm\fbox{\includegraphics[width=11.00cm]{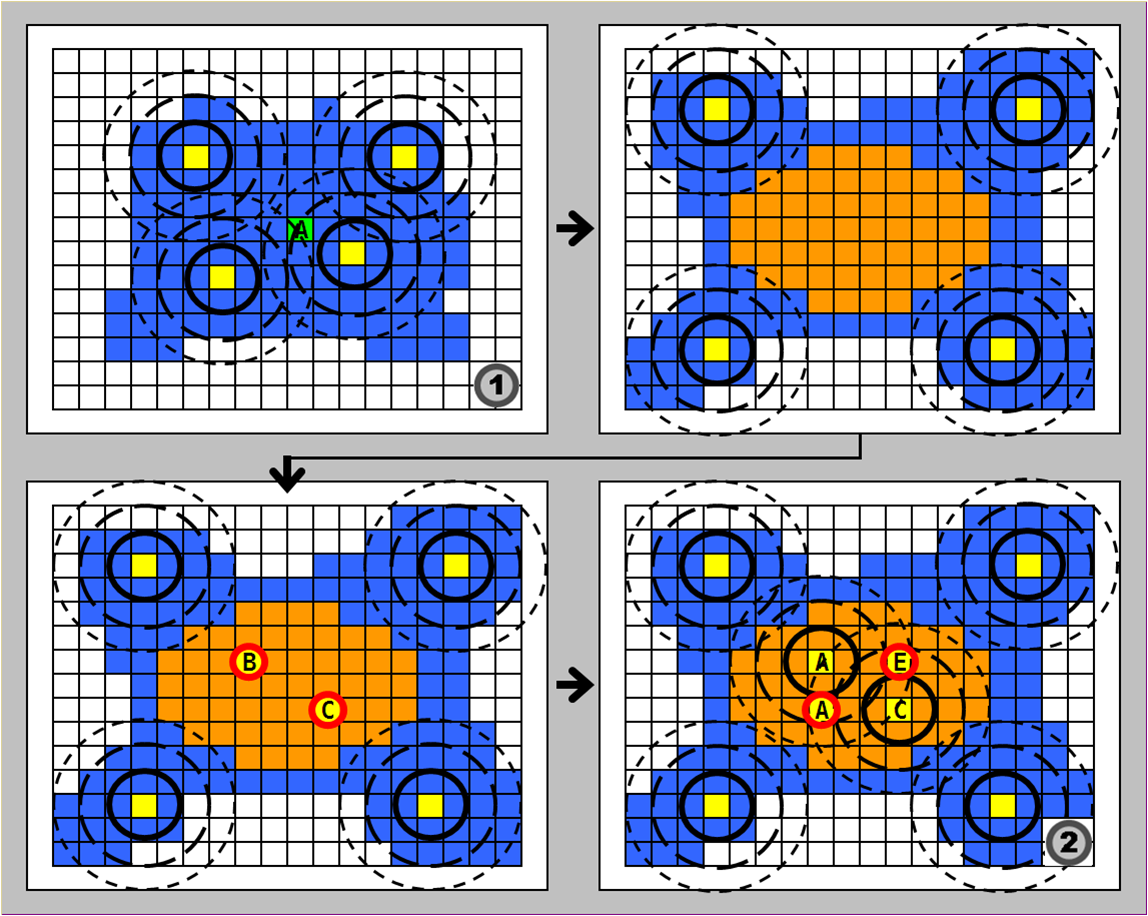}}}
\caption{Generation of new driver cells during a tumorigenic proliferation event. Thanks to a fault in the MOC-generation mechanism, oen or more daughter driver cells have the same MOC sequnece of the mother (A). Since this MOC sequence triggers a proliferation for the mother, the same holds true for the daughters.}
\label{figxx}
\end{center} \end{figure}

\begin{figure}[p] \begin{center}
{\fboxrule=0.2mm\fboxsep=0mm\fbox{\includegraphics[width=12.50cm]{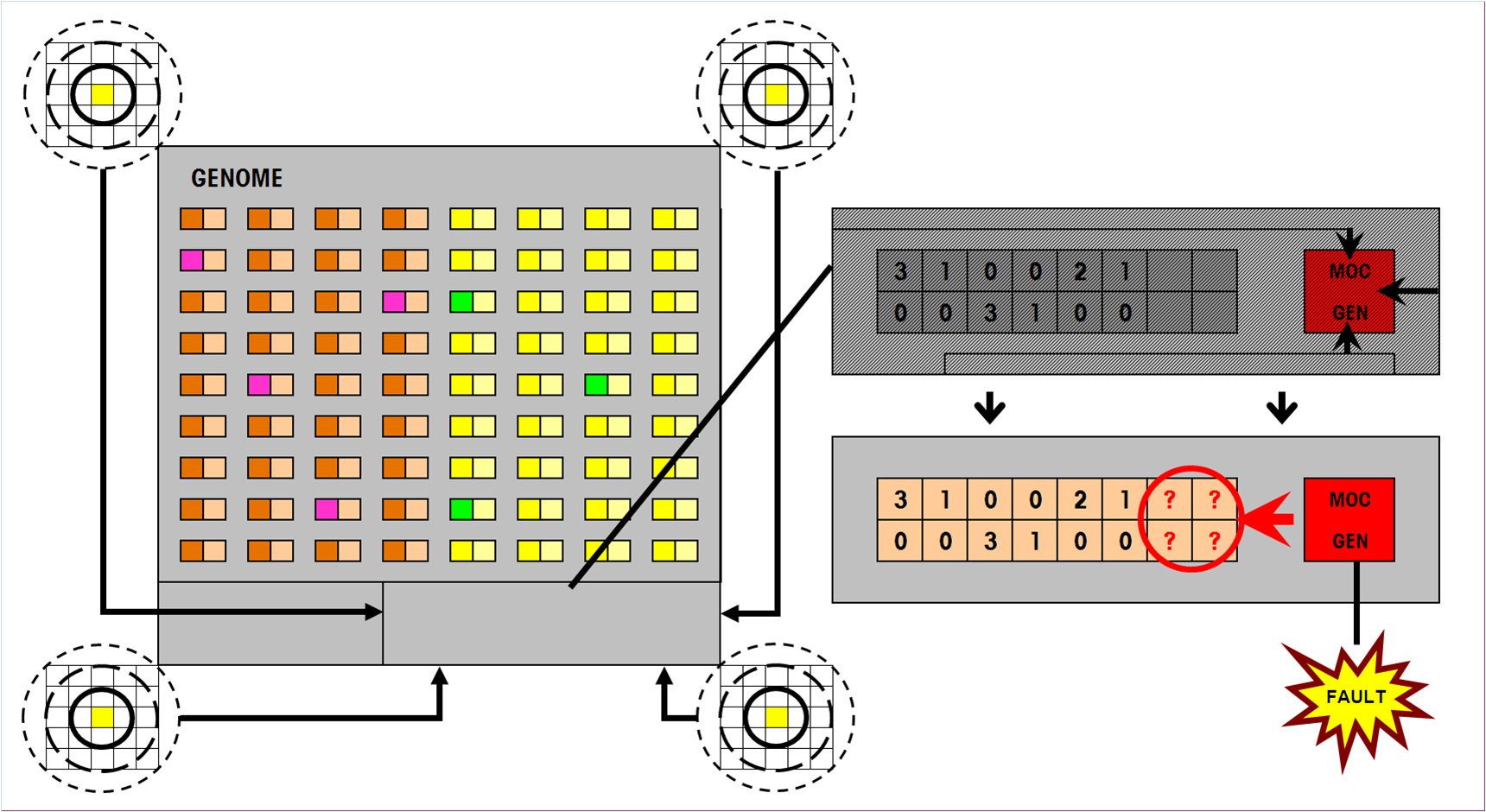}}}
\caption{A damage inside the red box impairs the differentiation machinery. It is hypothesised that the red box is implemented in nature by gene regulatory networks in which a key role is played by tumour-suppressor genes.}
\label{figxx}
\end{center} \end{figure}

\begin{figure}[p] \begin{center}
{\fboxrule=0.2mm\fboxsep=0mm\fbox{\includegraphics[width=13.50cm]{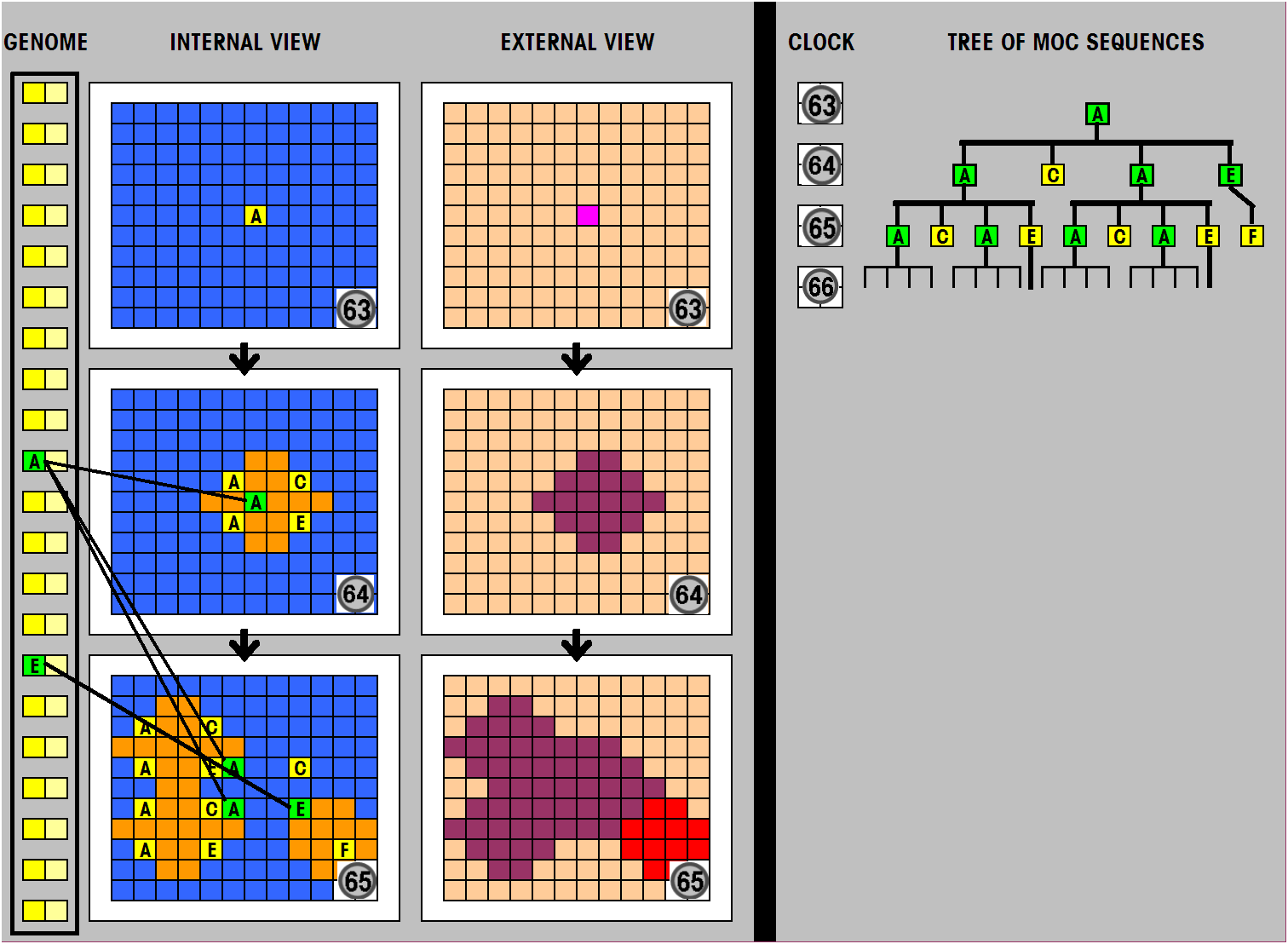}}}
\caption{Carcinogenic proliferation in a driver cell. A damage to the MOC generation mechanism of cell labelled with A has the effect of replacing daughter driver cells having MOC sequences B and D with additional driver cells having MOC sequence A, which in turn trigger other identical proliferation in the subsequent steps. In the tumour a hierarchy of cell types and/or cells with different degrees of differentiation arises. The amount of purple cells expands without limit.}
\label{figxx}
\end{center} \end{figure}

\clearpage

\pagebreak[4]

\section{Conclusions}

\subsection{A new approach to the study of biology}
 
In the so-called post-genomic era, it has become clear how the functioning of biological systems relies on hugely complicated networks of genes, proteins, chemical reactions. The most common approach to untertake the study of this complex matter can be described as ``divide et impera'' or ``bottom-up''. This translates to a narrowing of the field of expertise of many researchers, who have a very detailed knowledge in specific domains, and is also reflected in the multiplication of scientific journals specialised in specific sectors and sub-disciplines. Given the complexity of the subject, this approach has indeed been considered as the only possible one. Using a metaphor, we could say that biologists are trying to put together a huge puzzle, made up of hundreds of thousands of individual pieces, being each biologist or research laboratory concerned with a small part of the puzzle only.     
 
Another consequence of the complexity of biological systems is the enormous flow of biolological data produced by modern analysis techniques. In this context, the study of biology has become increasingly dominated by computer science. Examples of this trend are represented by the use of hardware and software tools for the sequencing of genomes of the most diverse species, the employment of specific software programmes for the automatic identification of gene sequences, the use of techniques of statistical analysis for the search of regular patterns in base sequences within genes, amino acid sequences within proteins, etc. The application of computer science tools to the analysis of biological data has given origin to a specific discipline called Bioinformatics. 
 
One reason why the ``bottom-up'' approach has been considered the only viable one is that biological systems are believed to be inherently more complex than any other systems present in nature or man-made. Mechanical systems, which are effectively described by means of mathematical equations, are by comparison much simpler. On the other hand, we can imagine what the results would be is one tried to apply the same ``puzzle'' approach to the study of, say, statistical mechanics. In physical systems an astonishing number of sub-atomic particles is present and one such approach would mean to study the properties of individual particles in the hope to understand the functioning of the system at the macroscopic level. If such an approach would have been employed, the study of statistical mechanics would have faced difficulties similar to those presently encountered by modern biology.      
 
``Epigenetic Tracking'', the model described in this work, has been constructed with a very different approach, that can be defined ``top-down'': first, we have drawn the general architecture of the model at a high level, on the base of known biological elements; subsequently, we have added additional elements -not necessarily known- in order to ``make things work in silico'' (i.e. to produce interesting behaviours through computer simulations); finally, we have come back to biology, trying to guess which real biological elements play the role of the additional elements. As a consequence the model may (and does) contain elements not necessarily adherent to current knowledge, but which can become a suggestion for biologists to look into new, previously unexplored directions.

\subsection{Experiments to prove the theory}
 
The effectiveness of the evo-devo method in generating shapes can be reconducted to some key features of the cellular model. The most important feature of the model is the presence of two categories of cells: normal cells and driver cells. Driver cells represent the scaffold of the embryo and make it possible to steer development by acting on a small subset of cells. The MOC takes different values in different driver cells and represents the source of differentiation during development, leading different driver cells at different times to trigger the activation of different developmental genes. This feature represents a key difference with respect to most other cellular models that rely on the positional information and the chemical micro-environment as basic providers of the information necessary for differentiation. 
 
From what we said, it appears clear that the first step towards proving this theory is the identification of natural driver cells. According to Epigenetic Tracking, cells tend spontaneously to form a hierarchy, made up of normal cells and driver cells. Whenever a cellular proliferation takes place, new driver cells are created so that a uniform distribution of driver cells is always maintained in the cellular system. The mechanism to create new driver cells is hypothesised to rely on the diffusion of chemical messengers from existing driver cells. Experimental efforts should aim at identifying such messengers and at reconstructing the process of generation of driver cells. 
 
In a series of experiments conducted a century ago, Hans Spemann demonstrated the potential of an area of the embryo, when transplanted into a second embryo, to induce the formation of a specific structure (the notochord) in an ectopic position. This area, called by Spemann ``organiser'', has ever since been referred to as ``Spemann's organiser''. The explanation of such experimental evidence by Epigenetic Tracking is straightforward: the transplanted area contains a driver cell which is the precursor of the notochord. The experiment of Spemann should be repeated, with the objective to identify such driver cell, keeping in mind that the generation of new driver cells is a dynamic process. The theory described foresees that natural driver cells are present in all organs of an organism's body, for the entire duration of the organism's life. Once natural driver cells have been correlated to specific biochemical markers, experiments should be aimed at finding such cells in the adult.
 
As we have seen, Germline Penetration implements a flow of genetic information from somatic cells to germline cells, to be passed on to future generations in a non-Mendelian fashion: as such, it can be considered the carrier of a transposon-mediated inheritance device. We mentioned how the biological phenomenon of sperm-mediated gene transfer could indeed represent the mechanism used by nature to implement the final stage of the Germline Penetration procedure, namely the phase in which MOC sequences enter the germline genome and copy themselves into MOS sequences of developmental genes. The next logical step would be to try to verify the first part of the process, the one in which the transposable-elements exit the driver cells in which they are hosted to reach the circulatory system and make their way towards germline cells.
 
As we mentioned, while the experimental evidence reported in \cite{EB10OG} suggests that the spreading of new transposable-elements into the genome of a lineage and the occurrence of major changes in the lineage are simultaneous events, our model predicts that the spread of transposons is an event which immediately (in evolutionary terms) follows the evolutionary change. This prediction is somewhat counterintuive, as one would expect quite the opposite to be true. If the sensitivity of the analysis techniques used to estimate the age of DNA sequences were able to prove such prediction, this would represent a clear indication in favour of the proposed model. 

\subsection{Final remarks}

``Epigenetic Tracking'' is a model of systems of biological cells, able to generate arbitrary 2 or 3-dimensional cellular shapes of any kind and complexity (in terms of number of cells, number of colours, etc.) starting from a single cell. If the complexity of such structures is interpreted as a metaphor for the complexity of biological structures, we can conclude that this model has the potential to generate the complexity typical of living beings. It has been shown how the model is able to reproduce a simplified version of some key biological phenomena such as development, the presence of ``junk DNA'', the phenomenon of ageing and the process of carcinogenesis. This model links the properties and behaviour of genes and cells to the properties and behaviour of the organism, describing and interpreting the said phenomena with a unified framework: for this reason, we think it can be proposed as a model for all biology. Future work will be aimed at closing the gap with molecular biology, mapping the model variables to individual genes and chemical elements.

\pagebreak[4]

\bibliographystyle{plain}
\bibliography{ldanxmodab}
 
\end{document}